\documentclass[conference]{IEEEtran}

\AtBeginDocument{%
  \providecommand\BibTeX{{%
    \normalfont B\kern-0.5em{\scshape i\kern-0.25em b}\kern-0.8em\TeX}}}

\usepackage[noadjust]{cite}
\usepackage{fancyhdr}
\usepackage{amsmath,amsfonts}
\usepackage{algorithmic}
\usepackage[ruled,linesnumbered,vlined]{algorithm2e}
\usepackage{graphicx}
\usepackage{textcomp}
\usepackage{xcolor}
\usepackage{url}
\usepackage{xspace}
\usepackage{balance}
\usepackage{booktabs, multirow} %
\usepackage{listings}

\usepackage{amsthm}
\newtheorem{definition}{Definition}
\usepackage{booktabs}
\usepackage{tikz}
\usepackage{wrapfig}
\usetikzlibrary{tikzmark}
\usepackage{makecell}
\usepackage{multicol}

\usepackage{hyperref}  %
\hypersetup{
hidelinks,
colorlinks=true,
linkcolor=teal,
citecolor=teal,
}
\usepackage{subcaption}
\begin{document}
\newcommand{\precaptionspace}{\vspace{-3ex}}
\newcommand{\pretabcaptionspace}{\vspace*{-1.5ex}}
\newcommand{\posttabcaptionspace}{\vspace*{-1.5ex}}
\newcommand{\Comment}[1]{}
\newcommand{\Space}[1]{}
\newcommand{\parabf}[1]{\noindent\textbf{#1}}
\newcommand{\smallspace}{\hspace{0.5em}}
\newcommand{\ccc}[1]{\hfill \begin{small}\# #1\end{small}}
\newcommand{\Num}[1]{\textbf{#1}}
\newcommand{\CodeIn}[1]{\begin{small}\texttt{#1}\end{small}}
\newcommand{\Fix}[1]{{\color{red}\bfseries [#1]}}
\newcommand{\lingming}[1]{{\color{red}\bfseries [Lingming: #1]}}
\newcommand{\yinlin}[1]{{\color{olive}\itshape [Yinlin: #1]}}
\newcommand{\chenyuan}[1]{{\color{teal}\itshape [Chenyuan: #1]}}
\newcommand{\jiayi}[1]{{\color{blue}\itshape [Jiayi: #1]}}
\newcommand{\hanchi}[1]{{\color{purple}\itshape [Hanchi: #1]}}
\newcommand{\yuxing}[1]{{\color{brown}\itshape [Yuxing: #1]}}
\newcommand{\jiawei}[1]{{\color{magenta}\itshape [Jiawei: #1]}}

\newcommand{\tech}{$\nabla$Fuzz\xspace} %
\newcommand{\pt}{PyTorch\xspace}
\newcommand{\tf}{TensorFlow\xspace}
\newcommand{\jax}{JAX\xspace}
\newcommand{\oneflow}{OneFlow\xspace}
\newcommand{\keras}{Keras\xspace}

\newcommand{\python}{Python\xspace}
\newcommand{\cpp}{C++\xspace}

\newcommand{\freefuzz}{FreeFuzz\xspace}
\newcommand{\docter}{DocTer\xspace}
\newcommand{\deeprel}{DeepREL\xspace}

\newcommand{\cradle}{CRADLE\xspace}
\newcommand{\lemon}{LEMON\xspace}
\newcommand{\predoo}{Predoo\xspace}
\newcommand{\audee}{AUDEE\xspace}
\newcommand{\muffin}{Muffin\xspace}
\newcommand{\grist}{GRIST\xspace}
\newcommand{\eagle}{EAGLE\xspace}
\newcommand{\nnsmith}{NNSmith\xspace}

\let\oldnl\nl%
\newcommand{\nonl}{\renewcommand{\nl}{\let\nl\oldnl}}%

\newcommand{\lbracket}{\langle}
\newcommand{\rbracket}{\rangle}

\newcommand{\OR}{\textbf{ or }}
\newcommand{\AND}{\textbf{ and }}
\newcommand{\NOT}{\textbf{ not }}

\newcommand{\ad}{AD\xspace}
\newcommand{\nd}{ND\xspace}

\newcommand{\jacobian}{Jacobian\xspace}
\newcommand{\hessian}{Hessian\xspace}

\newcommand{\numNeighbor}{5\xspace}

\newcommand{\numPtTotalAPI}{1592\xspace}
\newcommand{\numPtCoverAPI}{1071\xspace}
\newcommand{\numPtLoc}{1936K\xspace}
\newcommand{\numPtStars}{57.7K\xspace}
\newcommand{\verPt}{1.11\xspace}

\newcommand{\numPtTotalBugs}{80\xspace}
\newcommand{\numPtConfirmedBugs}{77\xspace}
\newcommand{\numPtUnknownBugs}{62\xspace}
\newcommand{\numPtKnownBugs}{15\xspace}
\newcommand{\numPtFixBugs}{18\xspace}
\newcommand{\numPtRejectBugs}{3\xspace}

\newcommand{\numPtDirect}{11\xspace}
\newcommand{\numPtAD}{64\xspace}
\newcommand{\numPtRev}{33\xspace}
\newcommand{\numPtFwd}{9\xspace}
\newcommand{\numPtND}{2\xspace}

\newcommand{\numPtOutput}{31\xspace}
\newcommand{\numPtFirstGrad}{44\xspace}
\newcommand{\numPtSecondGrad}{2\xspace}
\newcommand{\numPtGrad}{46\xspace}

\newcommand{\ptRevAPI}{torch.autograd.grad\xspace}
\newcommand{\ptFwdAPI}{torch.autograd.forward\_ad\xspace}
\newcommand{\ptNDAPI}{torch.autograd.gradcheck\xspace}

\newcommand{\ptCodeRevAPI}{\CodeIn{\ptRevAPI}\xspace}
\newcommand{\ptCodeFwdAPI}{\CodeIn{\ptFwdAPI}\xspace}
\newcommand{\ptCodeNDAPI}{\CodeIn{\ptNDAPI}\xspace}

\newcommand{\ptTimeCost}{25.7h\space}
\newcommand{\freefuzzPtTimeCost}{3.1h\xspace}

\newcommand{\numTFTotalAPI}{6381\xspace}
\newcommand{\numTFCoverAPI}{1902\xspace}
\newcommand{\numTFLoc}{2788K\xspace}
\newcommand{\numTFStars}{167K\xspace}
\newcommand{\verTF}{2.9\xspace}

\newcommand{\numTFTotalBugs}{29\xspace}
\newcommand{\numTFConfirmedBugs}{23\xspace}
\newcommand{\numTFUnknownBugs}{18\xspace}
\newcommand{\numTFKnownBugs}{5\xspace}
\newcommand{\numTFFixBugs}{2\xspace}
\newcommand{\numTFRejectBugs}{2\xspace}

\newcommand{\numMuffinCoverAPI}{79\xspace}

\newcommand{\numTFConfirmedBugsMuffinBudget}{10\xspace}
\newcommand{\numTFMuffinCanFindBugs}{0\xspace}
\newcommand{\numTFFreeFuzzCanFindBugs}{3\xspace}

\newcommand{\numTFDirect}{3\xspace}
\newcommand{\numTFAD}{18\xspace}
\newcommand{\numTFRev}{5\xspace}
\newcommand{\numTFFwd}{4\xspace}
\newcommand{\numTFND}{2\xspace}

\newcommand{\numTFOutput}{4\xspace}
\newcommand{\numTFFirstGrad}{17\xspace}
\newcommand{\numTFSecondGrad}{2\xspace}
\newcommand{\numTFGrad}{19\xspace}

\newcommand{\tfRevAPI}{tf.GradientTape.gradient\xspace}
\newcommand{\tfFwdAPI}{tf.autodiff.ForwardAccumulator\xspace}
\newcommand{\tfNDAPI}{tf.test.compute\_gradient\xspace}

\newcommand{\tfCodeRevAPI}{\CodeIn{\tfRevAPI}\xspace}
\newcommand{\tfCodeFwdAPI}{\CodeIn{\tfFwdAPI}\xspace}
\newcommand{\tfCodeNDAPI}{\CodeIn{\tfNDAPI}\xspace}

\newcommand{\tfTimeCost}{24.3h\space}
\newcommand{\tfTimeCostSeedOnly}{2.9h\space}
\newcommand{\freefuzzTfTimeCost}{3.9h\xspace}

\newcommand{\numOneflowTotalAPI}{409\xspace}
\newcommand{\numOneflowCoverAPI}{299\xspace}
\newcommand{\numOneflowLoc}{350K\xspace}
\newcommand{\numOneflowStars}{3.6K\xspace}
\newcommand{\verOneflow}{0.7.0\xspace}

\newcommand{\numOneflowTotalBugs}{30\xspace}
\newcommand{\numOneflowConfirmedBugs}{21\xspace}
\newcommand{\numOneflowUnknownBugs}{17\xspace}
\newcommand{\numOneflowKnownBugs}{4\xspace}
\newcommand{\numOneflowFixBugs}{10\xspace}
\newcommand{\numOneflowRejectBugs}{0\xspace}

\newcommand{\numOneflowDirect}{16\xspace}
\newcommand{\numOneflowAD}{5\xspace}
\newcommand{\numOneflowRev}{5\xspace}
\newcommand{\numOneflowFwd}{N/A\xspace}
\newcommand{\numOneflowND}{N/A\xspace}

\newcommand{\numOneflowOutput}{16\xspace}
\newcommand{\numOneflowFirstGrad}{2\xspace}
\newcommand{\numOneflowSecondGrad}{3\xspace}
\newcommand{\numOneflowGrad}{5\xspace}

\newcommand{\numOneflowTestFiles}{519\xspace}
\newcommand{\numOneflowDoc}{37\xspace}
\newcommand{\numOneflowModel}{51\xspace}

\newcommand{\oneflowRevAPI}{oneflow.autograd.grad\xspace}
\newcommand{\oneflowCodeRevAPI}{\CodeIn{\oneflowRevAPI}\xspace}

\newcommand{\oneflowTimeCost}{6.9h\space}

\newcommand{\numJaxTotalAPI}{791\xspace}
\newcommand{\numJaxCoverAPI}{634\xspace}
\newcommand{\numJaxLoc}{135K\xspace}
\newcommand{\numJaxStars}{19.7K\xspace}
\newcommand{\ratioJaxCover}{80.2\%\xspace}
\newcommand{\verJax}{0.3.14\xspace}

\newcommand{\numJaxTotalBugs}{34\xspace}
\newcommand{\numJaxConfirmedBugs}{23\xspace}
\newcommand{\numJaxUnknownBugs}{20\xspace}
\newcommand{\numJaxKnownBugs}{3\xspace}
\newcommand{\numJaxFixBugs}{7\xspace}
\newcommand{\numJaxRejectBugs}{1\xspace}

\newcommand{\numJaxDirect}{3\xspace}
\newcommand{\numJaxAD}{20\xspace}
\newcommand{\numJaxRev}{3\xspace}
\newcommand{\numJaxFwd}{1\xspace}
\newcommand{\numJaxND}{0\xspace}

\newcommand{\numJaxOutput}{14\xspace}
\newcommand{\numJaxFirstGrad}{8\xspace}
\newcommand{\numJaxSecondGrad}{1\xspace}
\newcommand{\numJaxGrad}{9\xspace}

\newcommand{\numJaxTestFiles}{83\xspace}

\newcommand{\jaxRevAPI}{jax.jacrev\xspace}
\newcommand{\jaxFwdAPI}{jax.jacfwd\xspace}
\newcommand{\jaxNDAPI}{jax.test\_util.check\_grads\xspace}

\newcommand{\jaxCodeRevAPI}{\CodeIn{\jaxRevAPI}\xspace}
\newcommand{\jaxCodeFwdAPI}{\CodeIn{\jaxFwdAPI}\xspace}
\newcommand{\jaxCodeNDAPI}{\CodeIn{\jaxNDAPI}\xspace}

\newcommand{\jaxTimeCost}{12.5h\space}

\newcommand{\numTotalBugs}{173\xspace}
\newcommand{\numConfirmedBugs}{144\xspace}
\newcommand{\numUnknownBugs}{117\xspace}
\newcommand{\numKnownBugs}{27\xspace}
\newcommand{\numFixBugs}{38\xspace}
\newcommand{\numRejectBugs}{6\xspace}

\newcommand{\numDirect}{33\xspace}
\newcommand{\numAD}{107\xspace}
\newcommand{\numRev}{46\xspace}
\newcommand{\numFwd}{14\xspace}
\newcommand{\numADMutual}{47\xspace} %
\newcommand{\numND}{4\xspace}

\newcommand{\numOutput}{65\xspace}
\newcommand{\numFirstGrad}{71\xspace}
\newcommand{\numSecondGrad}{8\xspace}
\newcommand{\numGrad}{79\xspace}

\newcommand{\freefuzzDetect}{21\xspace}

\newcommand{\ptOutputFPR}{19.3\%\xspace}
\newcommand{\tfOutputFPR}{8.3\%\xspace}
\newcommand{\jaxOutputFPR}{11.1\%\xspace}
\newcommand{\oneflowOutputFPR}{12.5\%\xspace}
\newcommand{\OutputFPR}{15.0\%\xspace}

\newcommand{\ptGradFPR}{21.2\%\xspace}
\newcommand{\tfGradFPR}{21.1\%\xspace}
\newcommand{\jaxGradFPR}{21.0\%\xspace}
\newcommand{\oneflowGradFPR}{25.0\%\xspace}
\newcommand{\GradFPR}{22.4\%\xspace}

\newcommand{\ptGradDiffFPR}{25.5\%\xspace}
\newcommand{\tfGradDiffFPR}{34.8\%\xspace}
\newcommand{\jaxGradDiffFPR}{58.1\%\xspace}
\newcommand{\oneflowGradDiffFPR}{25.0\%\xspace}
\newcommand{\GradDiffFPR}{37.7\%\xspace}

\newcommand{\ptGradPreFPR}{57.3\%\xspace}
\newcommand{\tfGradPreFPR}{46.4\%\xspace}
\newcommand{\jaxGradPreFPR}{68.6\%\xspace}
\newcommand{\oneflowGradPreFPR}{64.0\%\xspace}
\newcommand{\GradPreFPR}{59.8\%\xspace}

\newcommand{\ptGradNAFPR}{61.9\%\xspace}
\newcommand{\tfGradNAFPR}{53.1\%\xspace}
\newcommand{\jaxGradNAFPR}{78.2\%\xspace}
\newcommand{\oneflowGradNAFPR}{64.0\%\xspace}
\newcommand{\GradNAFPR}{66.7\%\xspace} %

\newcommand{\ptFPR}{20.7\%\xspace}
\newcommand{\tfFPR}{16.1\%\xspace}
\newcommand{\jaxFPR}{17.3\%\xspace}
\newcommand{\oneflowFPR}{20.0\%\xspace}
\newcommand{\FPR}{19.3\%\xspace}

\newcommand{\pointx}{\boldsymbol{x}\xspace}
\newcommand{\outputy}{\boldsymbol{y}\xspace}
\newcommand{\gradf}{\mathbf{G}\xspace}
\newcommand{\mR}{\mathbb{R}\xspace}

\newcommand{\tanU}{\mathbf{u}\xspace}
\newcommand{\cotanV}{\mathbf{v}\xspace}
\newcommand{\vectorF}{\mathbf{f}\xspace}
\newcommand{\scalarF}{f\xspace}
\newcommand{\gF}{\mathbf{f}'\xspace}

\newcommand{\jvp}{JVP\xspace}
\newcommand{\vjp}{VJP\xspace}

\newcommand\scalemath[2]{\scalebox{#1}{\mbox{\ensuremath{\displaystyle #2}}}}

\newcommand{\revision}[1]{{{#1}}}

\title{Fuzzing Automatic Differentiation in Deep-Learning Libraries}%

\captionsetup[figure]{font=bf,skip=0.2em} %
\captionsetup[table]{font=bf,skip=0.2em} %
\newcommand{\distance}{0.5em}
\setlength{\textfloatsep}{\distance} %
\setlength{\floatsep}{\distance} %
\setlength{\intextsep}{\distance} %
\setlength{\dbltextfloatsep}{\distance} %
\setlength{\dblfloatsep}{\distance} %

\author{
\IEEEauthorblockN{Chenyuan Yang}
\IEEEauthorblockA{University of Illinois \\Urbana-Champaign\\
 cy54@illinois.edu}
\\
\IEEEauthorblockN{Yuxing Tu}
\IEEEauthorblockA{Huazhong University of \\Science and Technology\\
yxtu@hust.edu.cn}

\and
\IEEEauthorblockN{Yinlin Deng}
\IEEEauthorblockA{University of Illinois \\Urbana-Champaign\\
yinlind2@illinois.edu}
\\
\IEEEauthorblockN{Hanchi Li}
\IEEEauthorblockA{University of Science \\and Technology of China\\
slxiaochi@mail.ustc.edu.cn}

\and
\IEEEauthorblockN{Jiayi Yao}
\IEEEauthorblockA{The Chinese University of \\Hong Kong, Shenzhen\\
jiayiyao@link.cuhk.edu.cn}
\\
\IEEEauthorblockN{Lingming Zhang}
\IEEEauthorblockA{University of Illinois \\Urbana-Champaign\\
lingming@illinois.edu}
}

\maketitle

\pagestyle{fancy}
\cfoot{\thepage}
\renewcommand{\headrulewidth}{0pt}
\renewcommand{\footrulewidth}{0pt}

\begin{abstract}
Deep learning (DL) has attracted wide attention and has been widely deployed in recent years. 
As a result, more and more research efforts have been dedicated to testing DL libraries and frameworks. 
However, existing work largely overlooked one crucial component of any DL system, automatic differentiation (\ad), which is the basis for the recent development of DL. 
To this end, we propose \tech, the first general and practical approach
specifically targeting the critical AD component in DL libraries.
Our key insight is that each DL library API can be abstracted into a function processing tensors/vectors, which can be differentially tested under various execution scenarios (for computing outputs/gradients with different implementations).
We have implemented \tech as a fully automated API-level fuzzer targeting \ad in DL libraries, which utilizes differential testing
on different execution scenarios to test both first-order and high-order gradients, and also includes automated filtering strategies to
remove false positives caused by numerical instability. 
We have performed an extensive study on four of the most popular and actively-maintained DL libraries, \pt, \tf, \jax, and \oneflow.
The result shows that \tech substantially outperforms state-of-the-art fuzzers in terms of both code coverage and bug detection. 
To date, \tech has detected \numTotalBugs bugs for the studied DL libraries, with \numConfirmedBugs already confirmed by developers (\numUnknownBugs of which are previously unknown bugs and \numAD are related to \ad).
\revision{Remarkably, \tech contributed 58.3\% (7/12) of all high-priority \ad bugs for \pt and \jax during a two-month period.}%
None of the confirmed \ad bugs were detected by existing fuzzers.

\end{abstract}

\section{Introduction}
\label{sec:intro}

Recent years have witnessed the rapid advancement of deep learning (DL) research and the wide adoption of DL solutions/technologies in various application domains, e.g., natural language processing~\cite{young2018recent}, healthcare~\cite{esteva2019guide}, scientific discovery~\cite{AlphaFold2021}, and software engineering~\cite{goffi2016automatic,gu2018deep, li2019deepfl, li2020dlfix, xia2023repairstudy, xia2022alpharepair}. As a result, there is a growing concern about the correctness and reliability of such systems. 
For example, for a safety-critical application domain such as autonomous driving, a bug in the DL system can cause serious consequences or even death~\cite{garcia2020comprehensive}.

\begin{figure}[t]
\includegraphics[keepaspectratio=true,width=1.0\columnwidth]{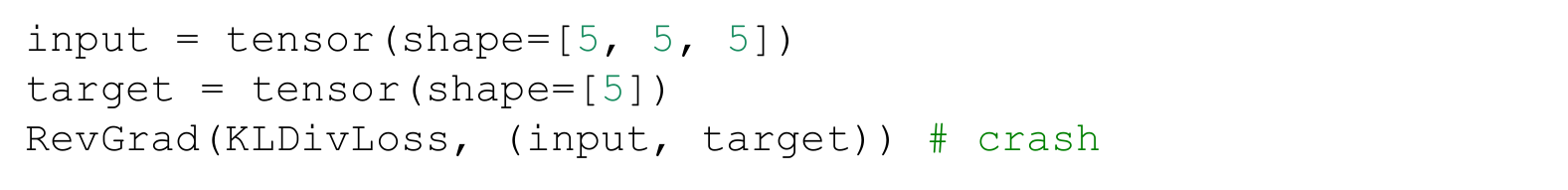}
\caption{Crash bug in \ad}
\label{fig:crash-bug}
\end{figure}

As it is critical to ensure the quality of increasingly influential DL systems, much research attention has been focused on testing/verifying DL models~\cite{shriver2021artifact,akhtar2018threat,carlini2019evaluating,gopinath2019symbolic,madry2017towards,moosavi2016deepfool,papernot2016limitations,pei2017deepxplore,grist,zhang2022towards,neuralattack2022} or application programs~\cite{cao2021characterizing, zhang2020detecting, lagouvardos2020static}. 
Recently, testing underlying DL libraries/frameworks (e.g., \pt/\tf) has also drawn wide attention, since DL libraries serve as the central infrastructure for all DL applications.%
\cradle~\cite{cradle} is one of the pioneering work to perform differential testing on multiple backends of \keras~\cite{keras} using various DL models.
\audee~\cite{audee} and \lemon~\cite{lemon} further apply search-based mutation strategies on existing models to generate more diverse test inputs.
While these mutation-based techniques heavily rely on seed models, \muffin~\cite{muffin} directly synthesizes DL models from DL APIs via a top-down generation approach.
Moreover, \muffin can detect inconsistencies in both the model training and inference phases across different backends of \keras. 
Unlike the above model-level testing techniques, the recent \freefuzz work~\cite{freefuzz} proposes a fully-automated API-level fuzzing technique via mining API inputs from open source.
Similarly, another recent work, \docter~\cite{docter}, directly generates inputs for each API based on DL-specific input constraints extracted from DL API documentation (assisted with human annotations). 
Despite the recent advances in DL library testing, existing techniques still suffer from a major limitation: 
The inference phase of DL models or the direct execution of DL APIs has received the most attention, while \emph{a crucial component of any DL system - automatic differentiation (\ad)~\cite{Merrienboer2018AutomaticDI} - is still understudied}. 
Many DL algorithms, notably back-propagation~\cite{enwiki:backprop}, one of the key algorithms for training feed-forward neural networks, rely heavily on \ad for derivative computation of arbitrary numerical functions.%
\ad enables the development of sophisticated DL models and algorithms since, without \ad, people would have to manually/symbolically calculate the derivatives of billions of parameters for large DL models~\cite{zhang2022opt}.
To obtain the derivatives automatically, special gradient/Jacobian computation operations in DL libraries need to be explicitly triggered (e.g., \CodeIn{tf.GradientTape.gradient()} in \tf).%
Notably, bugs in \ad may cause DL models to fail to converge and/or perform poorly in practical deployment, which is fatal for safety-critical applications.
For example, silent \ad computation bugs may cause the output of the deployed DL models to diverge significantly from the output in the training phase.
Besides, \ad bugs may also directly crash the entire training process, wasting massive computation resources when training recent popular large models and/or causing potential denial-of-service (DoS) attacks~\cite{dospaper}.
Figure~\ref{fig:crash-bug} shows a dangerous crash bug where a widely used PyTorch API \CodeIn{KLDivLoss}~\cite{kldiv_website_pt} will crash during \ad computation with a special input shape.%
However, such \ad engines have not been thoroughly tested by existing work.%

Although the recent \muffin work~\cite{muffin} can potentially test the training phase of DL models, it is still far from practical.
First, \muffin needs to \emph{manually} annotate the input constraints of considered DL APIs and use reshaping operations to ensure the validity of the generated models.%
As a result, \muffin can only cover a small set of APIs (confirmed in \S\ref{subsec:rq1}).
Second, whole model testing is inefficient, especially for large models/datasets.%
Third, false positives can originate from randomness, precision loss, and numerical instability, which are further amplified in the model training scenario.%
Fourth, differential testing of the training phase requires the same API interfaces across different DL libraries, which further limits its application. \muffin uses \keras and its supported backends \tf, Theano, and CNTK.
However, \keras 2.3.0 in 2019 is the last release supporting backends other than \tf~\cite{kerasmultibackend}.%
Lastly, \muffin cannot fully test the \ad engines in DL libraries, as it only covers part of reverse mode \ad and ignores forward mode \ad. 
In fact, the \muffin paper did not report any \emph{confirmed} \ad bug (confirmed in \S\ref{subsec:rq1}).

To thoroughly and automatically test the \ad engines in DL libraries, we propose \tech, the first general and practical framework specifically targeting the crucial \ad component. 
Our key insight is that each API in a DL library can be abstracted into a function processing tensors/vectors, which can be differentially tested under various execution scenarios (for computing outputs/gradients with different implementations). For example, the same DL API can be executed without \ad or with different \ad modes, but both the API output and gradient should be consistent across different execution scenarios, which can naturally serve as the oracle for differential testing.
In addition, since our test oracle is general at the function level, we can further transform each API into its gradient function to test the correctness of high-order gradient computation.

\tech can potentially address all the aforementioned limitations of \muffin. 
Through API-level testing, \tech no longer suffers from strict input constraints in model-level testing, and can be \emph{fully automated}. 
Also, API-level mutation is more efficient because it avoids extensive computation on large models with large datasets.
Besides, false positives can be significantly reduced, since floating-point precision loss will not be accumulated in API-level testing.%
Lastly, compared to \muffin, our technique is also more general, because we utilize the natural \ad oracles available in any DL library.

We have implemented \tech as a fully automated technique for API-level fuzzing with test oracles specifically targeting \ad in DL libraries. 
More precisely, while our approach is general and can leverage any existing API-level DL library fuzzer for input generation, we build \tech on top of state-of-the-art \freefuzz~\cite{freefuzzrepo} because it is fully automated and publicly available.
For test oracle, \tech automatically performs differential testing of each DL library API (and its high-order gradients) under different execution scenarios provided by the underlying DL library.
\tech also incorporates automated filter strategies to further reduce false positives caused by numerical instability issues.
We have conducted an extensive study of \tech on four of the most widely-used and actively-maintained DL libraries: \pt, \tf, \jax, and \oneflow. Our results show that \tech substantially outperforms state-of-the-art DL library fuzzers (including both \freefuzz and \muffin) in terms of both code coverage and bug detection. In fact, the bug in Figure~\ref{fig:crash-bug} is detected by \tech and cannot be detected by any previous techniques.%
Overall, our paper makes the following contributions:
\begin{enumerate}
    \item %
    To the best of our knowledge, this is the first work specifically targeting fuzzing the crucial \ad component in DL libraries with practical and general test oracles. %
    \revision{Our proposed \ad oracles can potentially strengthen and impact all future work on fuzzing DL libraries/systems.}

    \item  We have implemented \tech as a fully automated technique for testing \ad in DL libraries. 
    \tech is built on state-of-the-art \freefuzz and resolves the test oracle challenge with differential testing on various differentiation scenarios; \tech can also test the correctness of gradient computation of any order. Moreover, we have also designed novel strategies to filter out false positives.%
    \item  We conduct an extensive study on popular DL libraries (\pt, \tf, \jax, and \oneflow). \tech has detected \numTotalBugs bugs in total, with \numConfirmedBugs confirmed by developers (\numUnknownBugs are previously unknown and \numAD are \ad-related) and \numFixBugs already fixed. Remarkably, \tech contributed 58.3\% (7/12) of all high-priority \ad bugs for \pt and \jax within two months.
    None of the 107 \ad-related bugs can be detected by existing work. %
\end{enumerate}

\section{Background}
\label{sec:background}

\subsection{Basics about DL Libraries}
\begin{figure}
\centering
\includegraphics[keepaspectratio=true,width=\columnwidth]{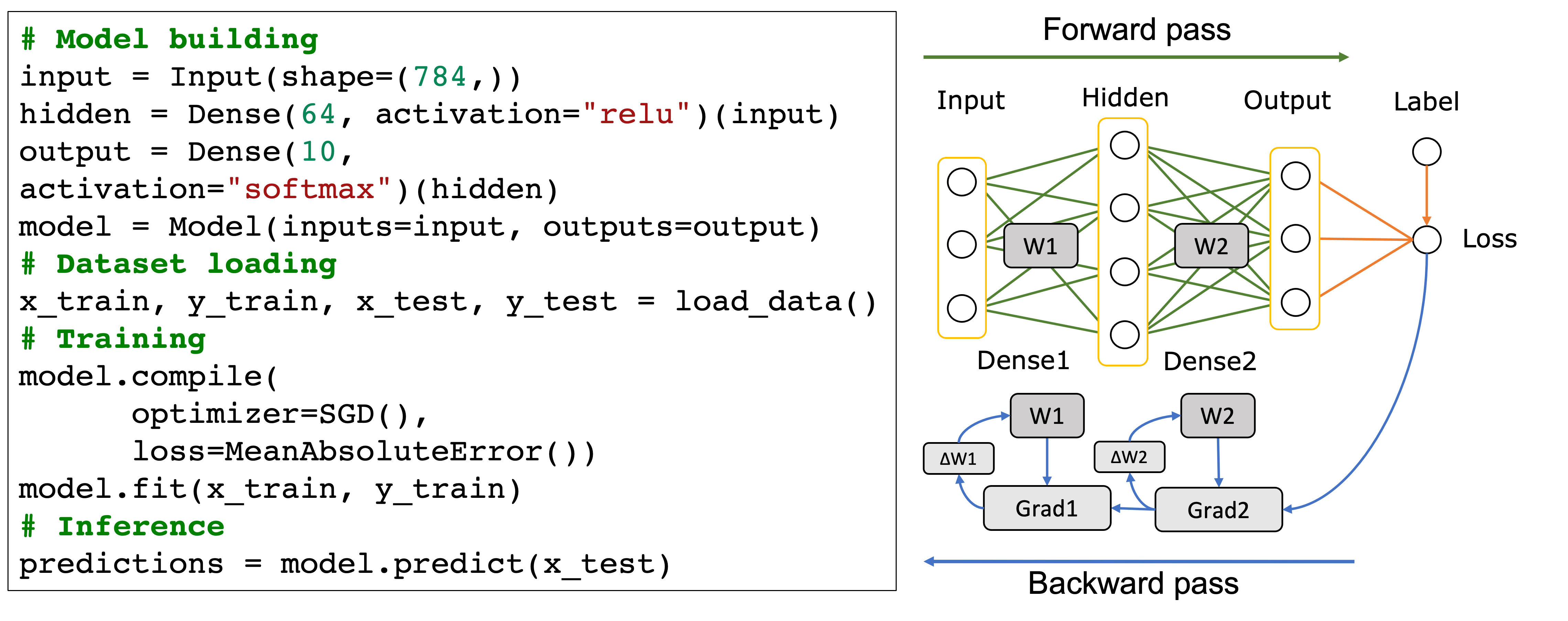}
\caption{An example of DL model training and inference}
\label{fig:tfmodel}
\end{figure}

\parabf{DL Models and DL APIs.} To develop a DL pipeline, users usually call DL APIs in a DL program to accomplish the following: build a DL model, load a dataset, train the DL model with labeled training data, and test it with evaluation data. 
The example \tf program shown in the left side of Figure~\ref{fig:tfmodel} constructs a dense neural network, which contains two \CodeIn{Dense} layers, with \CodeIn{relu} and \CodeIn{softmax} as the activation functions. 
Starting from an input layer \CodeIn{input}, the tensor \CodeIn{output} is obtained by invoking the DL APIs sequentially. 
The model is then constructed by defining the inputs and outputs, and compiled with specified optimizer (\CodeIn{SGD}) and loss function (\CodeIn{MeanAbsoluteError}). 
Next, we train the model with the high-level API \CodeIn{model.fit} and make predictions with \CodeIn{model.predict}.

\parabf{Training phase.} For model training, DL libraries usually provide high-level APIs (e.g., \CodeIn{model.fit()} in \tf) for ease of use. However, the actual training phase is complicated and composed of three stages: forward pass, loss computation, and backward pass. 
Such training steps will be carried out repeatedly until convergence.
Figure~\ref{fig:tfmodel} depicts an illustration of one training step for the example program, where \CodeIn{W1} and \CodeIn{W2} stand for the weight tensors of \CodeIn{Dense} layers. 
During the forward pass when the output tensor is computed with input and weight tensors, every executed operation will be automatically recorded for automatic differentiation (\ad). Please note that additional traced information is omitted in Figure~\ref{fig:tfmodel} for simplicity. 
After loss computation (requiring labels), the recorded \ad context will be used to compute gradients (\CodeIn{Grad1, Grad2}) of the loss w.r.t the weight tensors.
Lastly, the optimizer will apply the gradients to update \CodeIn{W1} and \CodeIn{W2} by adding \CodeIn{$\Delta$W1} and \CodeIn{$\Delta$W2} (computed from the gradients).

\parabf{Inference phase.} During the inference phase, DL APIs will be executed to compute the output tensor as in the forward pass, except that \ad is usually disabled for efficiency.

\subsection{Automatic Differentiation}

\label{sec:bg-ad}

Automatic differentiation (\ad) is one of the core components of DL frameworks, which contributes substantially to the success of DL.
\ad decomposes a function/model into a set of elementary operations for which derivatives are known and leverages chain rule to compose the derivatives of these operations~\cite{Baydin2017AutomaticDI}.
It allows us to calculate the derivative of any function/model without extensive manual effort.
\ad usually has two distinct modes, \textit{reverse mode} (or reverse accumulation) and \textit{forward mode} (or forward accumulation).

\newcommand{\xa}{x_1}
\newcommand{\xb}{x_2}
\newcommand{\va}{v_1}
\newcommand{\vb}{v_2}
\newcommand{\vc}{v_3}
\newcommand{\vd}{v_4}

\newcommand{\barxa}{\bar{x}_1}
\newcommand{\barxb}{\bar{x}_2}
\newcommand{\barva}{\bar{v}_1}
\newcommand{\barvb}{\bar{v}_2}
\newcommand{\barvc}{\bar{v}_3}
\newcommand{\barvd}{\bar{v}_4}
\newcommand{\barf}{\bar{f}}

\newcommand{\dotxa}{\dot{x}_1}
\newcommand{\dotxb}{\dot{x}_2}
\newcommand{\dotva}{\dot{v}_1}
\newcommand{\dotvb}{\dot{v}_2}
\newcommand{\dotvc}{\dot{v}_3}
\newcommand{\dotvd}{\dot{v}_4}

\begin{figure}[t]
\begin{subfigure}{0.34\columnwidth}
    \includegraphics[keepaspectratio=true,width=0.93\textwidth]{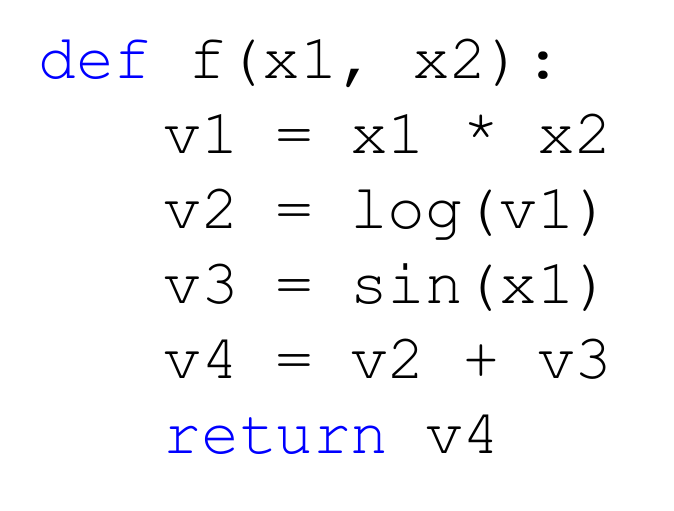}
\caption{Definition}
\label{fig:fndef}
\end{subfigure} 
\hspace{1em}
\begin{subfigure}{0.6\columnwidth}
\centering
\includegraphics[keepaspectratio=true,width=0.8\textwidth]{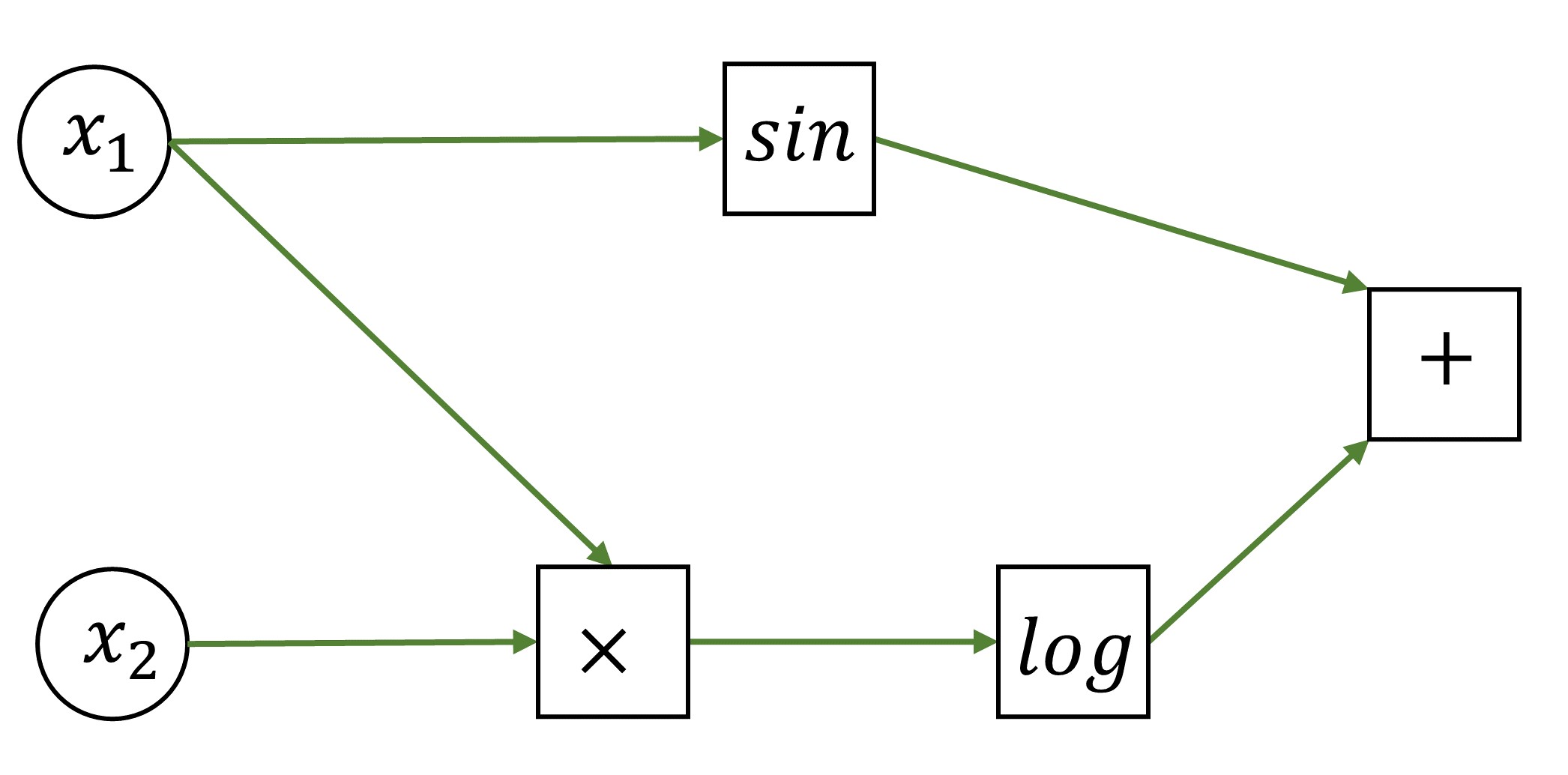}
\caption{Computational graph}
\label{fig:fngraph}
\end{subfigure}
\caption{Function $f(\xa,\xb) = \log(\xa\cdot \xb) + \sin(\xa)$}
\label{fig:examplefn}
\end{figure}

\parabf{Reverse Mode.}
Reverse mode is the most common \ad mode in DL libraries. 
It evaluates the chain rule from the output to the input, which is the reverse order of the original function/model.
Reverse mode calculates the derivative in two different phases: \textit{forward phase} and \textit{backward phase}. 
In the \textit{forward phase}, we will obtain the output of the original function, and evaluate the output value and all the intermediate variables, whose values are stored in memory. 
In the \textit{backward phase}, derivatives are calculated by leveraging the chain rule and the intermediate value, which could propagate back the derivative from the output to the input.

Figure~\ref{fig:examplefn} presents an example function $f(\xa,\xb)$ together with its computation graph.%
When $(\xa, \xb)$ is $(1, 2)$, the trace of computation of reverse mode AD is shown in Table~\ref{tab:revtrace}. 
To simplify the representations, we use $\bar{v} = \frac{\partial f}{\partial v}$ to represent the partial derivative of $f$ w.r.t the variant $v$. 
First, the function evaluates the output value ($f = 1.53$) and stores all the intermediate values in the forward phase, shown on the left side of Table~\ref{tab:revtrace}. 
Because the derivatives of the elementary operations are known, e.g., the derivative of $\sin(\xa)$ is $\cos(\xa)$, the reverse mode AD can leverage the chain rule and stored values to propagate back the derivative from $f$ to inputs $\xa, \xb$ automatically, shown on the right side of Table~\ref{tab:revtrace}. 

It is worth noting that derivatives $\bar{\xa}=1.54$ and $\bar{\xb}=0.5$ are calculated in just one reverse pass since the output value of this function is scalar. 
If the function is more general (e.g., $\vectorF: \mR^n\rightarrow \mR^m$), the reverse mode \ad needs $m$ reverse pass to calculate the gradients. 
As DL usually computes the gradient of a low dimensional tensor (e.g., scalar loss values) w.r.t a huge number of parameters in practice, reverse mode is the main \ad used in DL libraries. 

\begin{table}[!t]
\centering
\caption{Reverse mode AD computation trace}
\scalebox{0.9}{
\begin{tabular}{ll}
\toprule
\textbf{Forward Phase} & \textbf{Backward Phase} \\\midrule
\tikzmark{s1}$\xa = 1$ & \tikzmark{s2}$\barxa = 1.54$\\
$\xb = 2$ & $\barxb = 0.5$\\
\midrule
$\va = \xa \cdot \xb = 2$ & $\barxa = \barxa + \barva\cdot\frac{\partial \va}{\partial \xa} = 0.54 + 0.5\cdot2=1.54$\\
 & $\barxb = \barva\cdot\frac{\partial \va}{\partial \xb} = 0.5$\\
$\vb = \log(\va) = 0.69$ & $\barva = \barvb\cdot\frac{\partial \vb}{\partial \va} = 1 / \va = 0.5$\\
$\vc = \sin(\xa) = 0.84$ & $\barxa = \barvc\cdot\frac{\partial \vc}{\partial \xa} = \cos(1) = 0.54$\\
$\vd = \vb + \vc = 1.53$ & $\barvb = \barvd\cdot \frac{\partial \vd}{\partial \vb} = 1$\\
 & $\barvc = \barvd\cdot \frac{\partial \vd}{\partial \vc} = 1$\\
\midrule
\tikzmark{f1}$f = \vd = 1.53$ & \tikzmark{f2}$\barvd = \barf = \partial f / \partial \vd = 1$\\
\bottomrule
\end{tabular}}
\begin{tikzpicture}[overlay,remember picture]
\draw[->,thick] ([ shift={(-1.0ex,0ex)}]pic cs:s1) -- ([ shift={(-1.0ex,1.5ex)}] pic cs:f1);
\draw[->,thick] ([ shift={(-3.5ex,1.5ex)}]pic cs:f2) -- ([ shift={(-3.5ex,0ex)}] pic cs:s2);
\end{tikzpicture}
\label{tab:revtrace}
\end{table}

\parabf{Forward Mode.}
Different from reverse mode, forward mode \ad computes the derivatives simultaneously with the original function/model outputs, i.e., it evaluates the chain rule from input to output. 
Thus, it does not need to store the intermediate values like reverse mode. 
Forward mode \ad also has two phases: \textit{forward primal phase} and \textit{forward tangent phase}. 
The \textit{forward primal phase} obtains the output of the original function, while concurrently the \textit{forward tangent phase} calculates the gradient by applying the chain rule from input to output.%

Back to $f$ shown in Figure~\ref{fig:examplefn}, whose forward mode computational trace is shown in Table~\ref{tab:fwdtrace}. 
The forward mode AD computes the derivative by applying the chain rule to each elementary operation along the forward primal phase. 
However, it can only compute the gradient of one input in one pass. 
In this example, we calculate the partial gradient of $f$ w.r.t $\xa$. 
For simplicity, we define $\dot{v}=\frac{\partial v}{\partial \xa}$ as the partial gradient of $v$ w.r.t to the input $\xa$. 
Thus, we set $\dot{\xa} = 1, \dot{\xb}=0$ at the beginning since the gradient of $\xa$ w.r.t itself is 1 and $\xb$ does not affect $\xa$. 
Along the forward tangent phase shown on the right side of Table~\ref{tab:fwdtrace}, the final partial gradient $\dot{f} = \frac{\partial f}{\partial \xa}$ is 1.54, the same as computed by reverse mode.

In general, for function $\vectorF: \mR^n\rightarrow \mR^m$, forward mode \ad requires $n$ evaluations to calculate the gradient by setting $\dot{x_i}=1$ and the rest to zero for each input. 
Thus, forward mode \ad can be more time and memory efficient than reverse mode when $n \leq m$, and is useful in cases like computing the \hessian matrix~\cite{enwiki:hessian} efficiently.

\begin{table}[!t]
\centering
\caption{Forward mode AD computation trace}
\scalebox{0.9}{
\begin{tabular}{ll}
\toprule
\textbf{Forward Primal Phase} & \textbf{Forward Tangent Phase} \\\midrule
\tikzmark{s3}$\xa = 1$ & \tikzmark{s4}$\dotxa = 1$\\
$\xb = 2$ & $\dotxb = 0$\\
\midrule
$\va = \xa \cdot \xb = 2$ & $\dotva = \dotxa\cdot \xb = 2$\\
$\vb = \log(\va) = 0.69$ & $\dotvb = \dotva / \va = 1$\\
$\vc = \sin(\xa) = 0.84$ & $\dotvc = \dot{\xa}\cdot\cos(\xa) = 0.54$\\
$\vd = \vb + \vc = 1.53$ & $\dotvd = \dotvb + \dotvc = 1.54$\\
\midrule
\tikzmark{f3}$f = \vd = 1.53$ & \tikzmark{f4}$\dot{f} = \dotvd = 1.54$\\

\bottomrule
\end{tabular}}
\begin{tikzpicture}[overlay,remember picture]
\draw[->,thick] ([ shift={(-1.0ex,0ex)}]pic cs:s3) -- ([ shift={(-1.0ex,1ex)}] pic cs:f3);
\draw[->,thick] ([ shift={(-3.5ex,0ex)}]pic cs:s4) -- ([ shift={(-3.5ex,1ex)}] pic cs:f4);
\end{tikzpicture}
\label{tab:fwdtrace}
\end{table}

Despite the recent advances in DL library testing~\cite{freefuzz,muffin,lemon,cradle,audee,eagle,docter}, there is still limited work that can effectively test the crucial \ad component for DL libraries. 
Therefore, this paper aims to build the first practical fuzzing technique specifically targeting \ad in DL libraries.

\section{Preliminaries}
\label{sec:pre}

In this section, we will present the preliminaries for differentiation computations, which are essential for understanding \ad implementation in DL libraries and our approach.

\subsection{Mathematics behind Automatic Differentiation}

Differentiation is a process of computing the gradient for a given function at a given point. 
Especially, the gradient is defined for scalar-valued functions as below:

\begin{definition} 
\label{def:gradient}
\textbf{Gradient.} 
For a scalar-valued function $\scalarF:\mR^n\rightarrow \mR$ and a point $\pointx$, its gradient at $\pointx$ is defined as below:
\begin{equation}
    \nabla \scalarF(\pointx) = 
    \begin{bmatrix}
    \frac{\partial \scalarF}{\partial x_1} & \cdots & \frac{\partial \scalarF}{\partial x_n}
    \end{bmatrix}^T
\end{equation}
\end{definition}

The gradient can be further generalized to functions that return non-scalar values, namely the \jacobian matrix~\cite{enwiki:jacobian}:

\begin{definition} 
\label{def:jacobian}
\textbf{\jacobian.} 
The \jacobian matrix of a vector-valued function $\vectorF: \mR^n\rightarrow \mR^m$ is defined as an $m\times n$ matrix:

\begin{equation}
    \mathbf{J}(\vectorF) = 
    \begin{bmatrix}
    \nabla^T \scalarF_1\\
    \vdots \\
    \nabla^T \scalarF_m
    \end{bmatrix}
    =
    \begin{bmatrix}
    \frac{\partial \scalarF_1}{\partial x_1} & \cdots & \frac{\partial \scalarF_1}{\partial x_n} \\
    \vdots & \ddots & \vdots \\
    \frac{\partial \scalarF_m}{\partial x_1} & \cdots & \frac{\partial \scalarF_m}{\partial x_n}
    \end{bmatrix}
\end{equation}

where $\vectorF(\pointx) = (\scalarF_1(\pointx), \scalarF_2(\pointx), \dots, \scalarF_m(\pointx))$.
\end{definition}

In this paper, for simplicity, we use ``gradient'' to represent the \jacobian matrix for the vector-valued function. 
The gradient function of $\vectorF$ is defined as $\gF: \mR^n\rightarrow\mR^m\times\mR^n$, where $\gF(\pointx)$ is the \jacobian matrix of $\vectorF$ at the point $\pointx$. 

The gradient $\gF(\pointx)\in \mR^m\times\mR^n$ can also be considered as a linear map, which maps the tangent space~\cite{enwiki:tangent} of the domain of $\vectorF$ at the point $\pointx$ to the tangent space of the codomain of $\vectorF$ at the point $\vectorF(\pointx)$. Given this mapping of $\gF(\pointx): \mR^n \rightarrow \mR^m$, we can now define \jacobian-vector product (\jvp) and vector-\jacobian product (\vjp), which have been adopted as the theoretical basis for efficient DL training and implemented using forward-/reverse-mode \ad in DL libraries:

\begin{definition}
\label{def:jvp}
\parabf{\jacobian-vector product.}
Given an input point $\pointx\in \mR^n$ and a tangent vector $\tanU\in \mR^n$ from the tangent space of $\vectorF$ at $\pointx$, the \jacobian-vector product is defined as below:
\begin{equation}
    \mathrm{\jvp}(\pointx, \tanU) = \gF(\pointx)\cdot\tanU
\end{equation}
\end{definition}

\jvp computes the directional gradient, with direction $\tanU\in\mR^n$, for the function $\vectorF: \mR^n\rightarrow\mR^m$ at the point $\pointx\in\mR^n$, and is implemented with forward mode \ad in DL libraries. 
Back to the example shown in Figure~\ref{fig:examplefn}, the $(\dotxa, \dotxb)$ in the forward mode \ad trace (Table~\ref{tab:fwdtrace}) is the tangent vector $\tanU$ in the Definition~\ref{def:jvp}. 
As a result, for the input $\pointx\in\mR^n$ and tangent vector $\tanU\in\mR^n$, the forward mode \ad can compute \jvp in only one pass by setting $\dot{\pointx}=\tanU$. 
Meanwhile, computing the full \jacobian matrix requires $n$ passes with forward mode \ad.

\begin{definition}
\label{def:vjp}
\parabf{Vector-\jacobian product.}
Given an input point $\pointx\in \mR^n$ and a cotangent vector $\cotanV\in \mR^m$, the vector-\jacobian product is defined as the below mapping:
\begin{equation}
    \mathrm{\vjp}(\pointx, \cotanV) = \cotanV\cdot\gF(\pointx)
\end{equation}

\end{definition}

With direction $\cotanV\in \mR^m$, \vjp computes the adjoint directional gradient for the function $\vectorF: \mR^n\rightarrow\mR^m$ at the point $\pointx\in\mR^n$. 
Similarly, DL libraries implement the reverse mode \ad to compute \vjp. 
$\barf=1$ in the reverse mode \ad trace (Table~\ref{tab:revtrace}) is a special case of the cotangent vector $\cotanV$ when the output is scalar. 
Generally, for the input $\pointx\in\mR^n$ and cotangent vector $\cotanV\in\mR^m$ the reverse mode \ad is capable of calculating the \vjp in just one pass with initialization of $\bar{\vectorF} = \cotanV$. 
By contrast, it requires $m$ passes for reverse mode to compute the full \jacobian matrix.

\subsection{Numerical Differentiation}
\label{sec:bg-nd}

Numerical differentiation (\nd)~\cite{burden2015numerical} is another approach to estimating the derivatives of a function by using the values of the original function at some sampled points. 
The most common method is to use finite difference approximation. 
For example, for a scalar-valued function $\scalarF: \mR^n\rightarrow\mR$ at the point $\pointx$, we can calculate the partial derivative of $x_i$ by using \nd:
\begin{equation}
\label{eqn:numediff}
    \frac{\partial f(\boldsymbol{x})}{\partial x_i} \approx \frac{f(\boldsymbol{x}+ \epsilon\boldsymbol{e}_i) - f(\boldsymbol{x}-\epsilon\boldsymbol{e}_i)}{2\epsilon}
\end{equation}
where $\boldsymbol{e}_i$ is $i$-th unit vector and $\epsilon>0$ is a small step.

However, \nd can be inaccurate due to truncation and rounding errors~\cite{Baydin2017AutomaticDI}, especially for the low precision data type. 
Besides, the time cost of \nd is $O(n)$ for a gradient in $n$ dimensions, which is the primary barrier to its usage in DL library since $n$ can be as large as billions in DL models~\cite{zhang2022opt}. 
Therefore, DL libraries do not rely on \nd as the main approach to calculating the gradient. 
Instead, most DL libraries leverage \nd to cross-check their own implementations of gradient calculation during developer testing.%

In this work, we further augment \tech{} oracle with \nd. 
This is because two \ad modes may return the same wrong gradient, which cannot be detected by comparing reverse and forward modes (detailed in \S\ref{subsec:grad-check}). To our knowledge, we are also the first to adopt \nd for automated DL library fuzzing. 

\section{Approach}
\label{sec:approach}

\begin{figure*}[t]
    \captionsetup{justification=centering}
    \centering
    \includegraphics[keepaspectratio=true,width=0.8\textwidth]{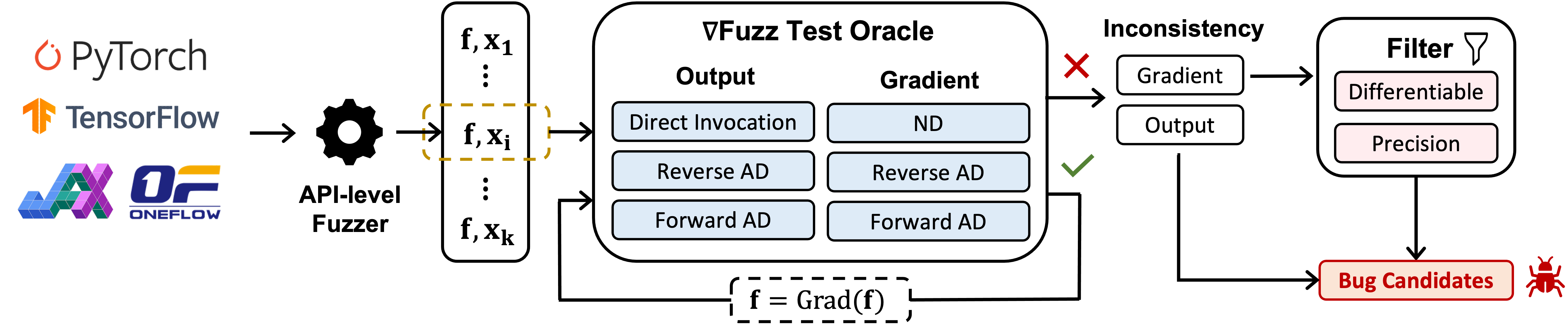}
    \caption{Overview of \tech}
    \label{fig:overview}
\end{figure*}

\newcommand{\outputDirect}{\outputy_{Direct}\xspace}
\newcommand{\outputRev}{\outputy_{Rev}\xspace}
\newcommand{\outputFwd}{\outputy_{Fwd}\xspace}

\newcommand{\gradRev}{\gradf_{Rev}\xspace}
\newcommand{\gradFwd}{\gradf_{Fwd}\xspace}
\newcommand{\gradND}{\gradf_{ND}\xspace}

Figure~\ref{fig:overview} shows the overview of our \tech approach for testing the \ad mechanism of DL libraries. 
Note that for ease of presentation, we abstract each DL library API under test into a function $\vectorF: \mR^n \rightarrow \mR^m$. \tech first invokes an off-the-shelf API-level fuzzer to generate input $\pointx \in \mR^n$ for the function (\S\ref{subsec:fuzzer}). 
Then \tech will cross-check its outputs and gradients at $\pointx$ in different execution scenarios (\S\ref{subsec:testoracle}). 
If $\vectorF$ passes the testing given $\pointx$ (without any inconsistency), \tech continues to test the higher-order gradient of $\vectorF$: \tech will wrap  $\vectorF$ to its gradient function $\vectorF': \mR^n \rightarrow \mR^m\times \mR^n$ and re-run the test oracle (\S\ref{subsec:highorder}). 
If there is any inconsistency (during first- or high-order gradient computation), \tech will filter out the false positives caused by numerical instability (\S\ref{subsec:filter}).
Finally, \tech returns the candidate bugs. 
The following sub-sections would explain each component in detail.

\subsection{API-level Fuzzer}
\label{subsec:fuzzer}

\tech's first component is an API-level fuzzer for generating inputs to invoke each DL API. Our approach is general, and can leverage any off-the-shelf API-level DL library fuzzer~\cite{freefuzz,docter}.
In this work, we leverage \freefuzz~\cite{freefuzz} to create the input for the function/API since it is fully automated and state-of-the-art. DL library APIs are often exposed in Python, a dynamically typed language, making it even hard to determine the input types for each DL API. To overcome this issue, \freefuzz automatically traces API inputs when executing code mined from various sources, including DL models, developer tests, and code snippets from DL documentation. \freefuzz further includes mutation strategies to generate more inputs based on the traced seed API inputs.  

DL library APIs may have configuration arguments (in addition to input tensors). 
For example, \CodeIn{torch.sum(input,dim)} returns the sum of each row of the \CodeIn{input} tensor in the dimension \CodeIn{dim}, which is a configuration argument. 
\freefuzz can generate inputs for both input tensors and configuration arguments. 
For each successful API invocation generated by \freefuzz, \tech would automatically create a wrapper function to transform the API invocation into a function mapping from the input to the output tensor(s).
Moreover, DL library APIs could take several multi-dimensional tensors as input/output, such as \CodeIn{tf.add(x,y)}, which adds two multi-dimensional tensors \CodeIn{x} and \CodeIn{y} element-wise. While \tech is directly applied to such APIs (with tensor input/output) in our implementation, we abstract each API into $\vectorF: \mR^n\rightarrow\mR^m$ for the ease of presentation. This abstraction can be viewed as flattening multi-dimensional tensors into vectors (and concatenating them if there are multiple input/output tensors).%

\subsection{Test Oracles}
\label{subsec:testoracle}

\SetKwData{revG}{revGrad}
\SetKwData{fwdG}{fwdGrad}
\SetKwData{ndgrad}{ndGrad}
\begin{algorithm}[t!]
\small
\caption{\tech oracle algorithm}
\label{alg:test}
\SetKwData{fn}{fn}
\SetKwData{input}{input}
\SetKwData{output}{output}
\SetKwData{status}{status}
\SetKwData{value}{value}
\SetKwData{statuses}{statuses}
\SetKwData{outputs}{outputs}
\SetKwData{values}{values}
\SetKwData{revS}{revStatus}
\SetKwData{revV}{revValue}
\SetKwData{revO}{revOutput}
\SetKwData{revT}{revTestResult}
\SetKwData{fwdS}{fwdStatus}
\SetKwData{fwdV}{fwdValue}
\SetKwData{fwdO}{fwdOutput}
\SetKwData{fwdT}{fwdTestResult}
\SetKwData{order}{order}
\SetKwData{curorder}{curOrder}
\SetKwFunction{Grad}{Grad}
\SetKwFunction{Test}{\tech-Oracle}
\SetKwFunction{CheckOutput}{IsOutputConsistent}
\SetKwFunction{Direct}{DirectInv}
\SetKwFunction{Fwd}{FwdInv}
\SetKwFunction{Rev}{RevInv}
\SetKwFunction{ND}{NDGrad}
\SetKwFunction{CheckGrad}{IsGradientConsistent}
\SetKwProg{Fn}{Function}{:}{}
\SetKwInOut{Input}{Input}
\SetKwInOut{Output}{Output}

\Fn{\Test{\fn, \input, \order}}{
  \Input{The function under test \fn, the function input \input, and the gradient order to be tested \order}
  \Output{The oracle outcome}
  \BlankLine
  \curorder $\leftarrow$ 1\label{lst:line:startline}\\
  \While{\curorder $\le$ \order\label{lst:line:rerun}}{
      \outputs $\leftarrow$ \Direct(\fn, \input, REP=10)\label{lst:line:directoutputs}\\
      \If{\NOT \CheckOutput(\outputs) \label{lst:line:checkdet}}{
        \algorithmicreturn{ RANDOM}
        \label{lst:line:retrandom}
      }
      \revO, \revG $\leftarrow$ \Rev(\fn, \input)\label{lst:line:rev}\\
      \fwdO, \fwdG $\leftarrow$ \Fwd(\fn, \input)\label{lst:line:fwd}\\
      
      \If{\NOT \CheckOutput(\outputs, \revO, \fwdO)\label{lst:line:check-output-inconsistent}}{
        \algorithmicreturn{ OUTPUT\_INCONSISTENT}
        \label{lst:line:output-inconsistent}
      }
      \ndgrad $\leftarrow$ \ND(\fn, \input)\\
      \If{\NOT \CheckGrad(\revG, \fwdG, \ndgrad) \label{lst:line:check-grad-inconsistent}}{
        \algorithmicreturn{ GRADIENT\_INCONSISTENT}
        \label{lst:line:grad-inconsistent}
      }
      \fn $\leftarrow$ \Grad(\fn)\label{lst:line:gradfn}\\
      \curorder $\leftarrow$ \curorder + 1
  }
  
  \algorithmicreturn{ PASS}
}

\end{algorithm}

\newcommand{\status}{status\xspace}
\newcommand{\statuses}{statuses\xspace}
\newcommand{\crash}{\texttt{Crash}\xspace}
\newcommand{\exception}{\texttt{Exception}\xspace}
\newcommand{\success}{\texttt{Success}\xspace}

As shown in Algorithm~\ref{alg:test}, the input to the \tech oracle algorithm is the function under test, an input for the function, and the highest order of gradient to be tested. 
We start with the first-order gradient test (Line~\ref{lst:line:startline}). 
\tech first checks the determinism of the function by directly invoking it with the given input for multiple (by default 10) times (Line~\ref{lst:line:directoutputs}). 
If the outputs are inconsistent, \tech will return RANDOM and terminate the fuzzing process for this function (Line~\ref{lst:line:checkdet}-\ref{lst:line:retrandom}). 
Otherwise, it continues to invoke this function with reverse- and forward-mode \ad (Line~\ref{lst:line:rev}-\ref{lst:line:fwd}). 
Then it compares outputs returned by direct invocation and invocations with \ad. 
If any inconsistency is detected (Line~\ref{lst:line:check-output-inconsistent}), \tech will skip the gradient check and return this output inconsistency (Line~\ref{lst:line:output-inconsistent}). 
Otherwise, \tech proceeds to check the correctness of gradient computation by comparing gradients calculated by reverse mode \ad, forward mode \ad, and \nd (Line~\ref{lst:line:check-grad-inconsistent}). 
It will return the inconsistency if these gradients are different (Line~\ref{lst:line:grad-inconsistent}).
If the function passes all the above checks and we want to keep testing the higher-order gradient computation (Line~\ref{lst:line:rerun}), \tech will transform the function to its gradient function (Line~\ref{lst:line:gradfn}). 
The main loop will continue to test this new function until the termination criterion is met, e.g., detecting inconsistency or passing the test for the highest-order gradient computation. We next present more details of our output and gradient checks.

\subsubsection{Output Check}
When calculating the gradient in reverse or forward mode \ad, some additional operations are always incurred, such as tracing or shape checking. 
Thus, the invocation with \ad may have different output from the direct invocation. 
However, the outputs in different execution scenarios should not differ, which means any inconsistency can potentially be a bug.
Therefore, \tech would compare the output of the direction invocation, as well as the invocations with reverse mode and forward mode \ad.

\begin{figure}[htb]
\includegraphics[keepaspectratio=true,width=1.0\columnwidth]{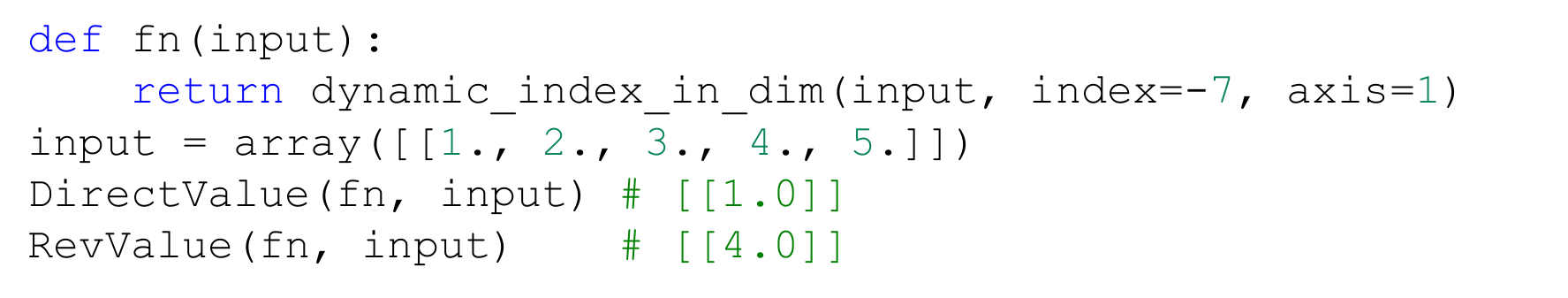}
\caption{Inconsistent outputs w/ and w/o \ad}
\label{fig:dynamic-bug}
\end{figure}

Take \jax API \CodeIn{jax.lax.dynamic\_index\_in\_dim} for instance, which performs integer indexing for input array~\cite{dynamic_index_website_jax}. 
Figure~\ref{fig:dynamic-bug} shows an example that its output values in direct invocation and reverse mode AD are different. 
The root cause of this issue is that reverse mode AD leads to the index being normalized multiple times, which changes the results for the out-of-bound negative index. 
This bug is detected by \tech and has been confirmed and fixed by the \jax developers.

\subsubsection{Gradient Check}
\label{subsec:grad-check}

Because reverse mode and forward mode apply different ways to calculate gradients, they could produce different gradients for the same function and input. Furthermore, \nd can be used to test the gradient computation of \ad since an improper formula may be adopted in both reverse and forward modes, resulting in the same incorrect gradient value.
Thus, \tech compares gradients computed by reverse
mode AD, forward mode AD, and ND to detect bugs.

For instance, a \pt API \CodeIn{torch.trace}~\cite{trace_website_pt} returns the sum of the diagonal elements of the input 2-D matrix. Obviously, an input with shape \CodeIn{(4,2)} has two elements in its diagonal, so this API will return the sum of these two elements. 
However, \textit{three} elements of the gradient computed in reverse mode have gradient 1, compared to only \textit{two} elements in forward mode.
This inconsistency is caused by the wrong formula used in reverse mode \ad for \CodeIn{torch.trace}. 
This bug found by \tech has been confirmed by the developers.

Here is another example showing the value of further leveraging \nd.
The \pt API \CodeIn{hardshrink(x,lambd)}~\cite{hardshrink_website_pt} returns \CodeIn{x} when \CodeIn{|x|}$>$\CodeIn{lambd}; otherwise, it just returns 0. 
That said, when \CodeIn{lambd} is 0, this API is equivalent to the linear function $y=x$.
However, it will have different gradients for input 0 in \ad and \nd with \CodeIn{lambd=0}. 
Both reverse and forward mode \ad return 0 as the gradient, while \nd returns 1 as the gradient. 
Obviously, the gradient should be 1.
This bug detected by \tech has also been confirmed in \pt.

Nevertheless, due to the drawbacks of \nd (e.g. truncating and rounding errors), it could produce inconsistent gradients. 
Thus, we only use \nd for the input with high precision, such as \CodeIn{float64}, which can minimize the effect of truncating and rounding errors.  
Furthermore, we design strategies to mitigate the false positives caused by the instability of \nd in \S\ref{subsec:filter}.

\subsubsection{High-order Gradients}
\label{subsec:highorder}
Besides the basic first-order gradient, \tech is capable of testing the correctness of higher-order gradient computation. 
To be more precise, \tech can take as input the gradient function $\vectorF':\mR^n\rightarrow\mR^m\times \mR^n$ of the current tested function $\vectorF: \mR^n\rightarrow\mR^m$ since our designed test oracles are general to the gradient function. Then, \tech can apply both output and gradient checks on $\vectorF'$. But how does the gradient for $\vectorF'$ (i.e., the second-order gradient for $\vectorF$) look like?
To illustrate it, let us first introduce \hessian matrix~\cite{enwiki:hessian}, the second-order gradient of scalar-valued functions:%

\begin{definition} 
\label{def:hessian}
\textbf{\hessian.} 
The \hessian matrix of a scalar-valued function $\scalarF: \mR^n\rightarrow \mR$ is defined as an $n\times n$ matrix as follows:

\begin{equation}
    \mathbf{H}(\scalarF) = 
    \begin{bmatrix}
    \frac{\partial^2 \scalarF}{\partial x_1^2} & \frac{\partial^2 \scalarF}{\partial x_1 \partial x_2} & \cdots & \frac{\partial^2 \scalarF}{\partial x_1\partial x_n} \\
    \frac{\partial^2 \scalarF}{\partial x_2\partial x_1} & \frac{\partial^2 \scalarF}{\partial x_2^2} & \cdots & \frac{\partial^2 \scalarF}{\partial x_2\partial x_n} \\
    \vdots & \vdots & \ddots & \vdots \\
    \frac{\partial^2 \scalarF}{\partial x_n\partial x_1} &\frac{\partial^2 \scalarF}{\partial x_n\partial x_2} & \cdots & \frac{\partial^2 \scalarF}{\partial x_n^2}
    \end{bmatrix}
\end{equation}
\end{definition}

In this way, the second-order gradient of a more general vector-valued function $\vectorF: \mR^n\rightarrow\mR^m$ can be defined as an $m\times n\times n$ matrix, which can be seen as an array of $m$ \hessian matrices.
Formally, for the function $\vectorF(\pointx) = (\scalarF_{i}(\pointx))_{i=1}^{m}$, the second-order gradient can be defined:
\begin{equation}
    \mathbf{H}(\vectorF) = (\mathbf{H}({\scalarF_1}), \mathbf{H}({\scalarF_2}), \dots, \mathbf{H}({\scalarF_m}))
\end{equation}

Notably, the current tested function can be the gradient of other functions. 
Hence, theoretically, \tech can test any order of gradient computation.
When the $\alpha$th-order gradient function passes all the testing described above and we want to test the correctness of its higher-order gradient computation, \tech would transform the $\alpha$th-order gradient function to its gradient function and re-run the test. 
In this work, we target the correctness of first- and second-order gradient computation since they are the most frequently used. 
Besides, to the best of our knowledge, very few existing DL libraries provide APIs calculating the gradient above the third order.

\begin{figure}[htb]
\includegraphics[keepaspectratio=true,width=1.0\columnwidth]{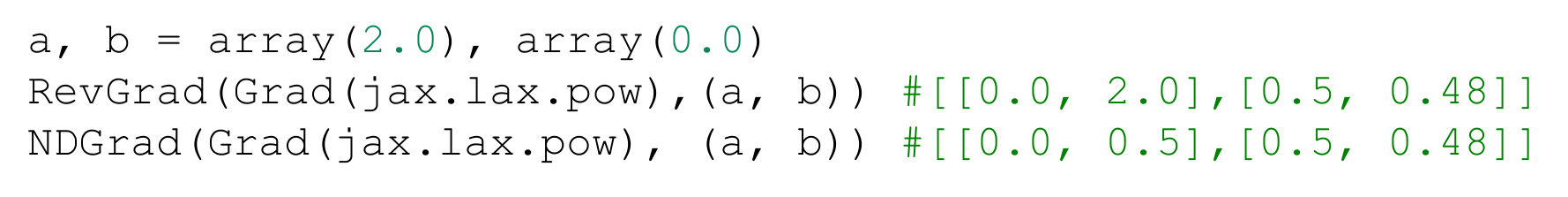}
	\caption{Inconsistent 2nd-order gradients in \ad and \nd}
	\label{fig:pow-bug}
\end{figure}

Take the \jax API \CodeIn{jax.lax.pow(a,b)}~\cite{pow_website_jax} for example, which returns $a^b$.
When the input \CodeIn{(a,b)} is \CodeIn{(2,0)}, this API can pass the test for the first-order gradient.
Then \tech will test the correctness of its second-order gradient computation given this input, which is shown in Figure~\ref{fig:pow-bug}. 
It turns out that the second-order gradient computed in reverse mode \ad is different than the one in \nd, while the latter is correct. 
This is because the gradient $\frac{\partial^2 f}{\partial a\partial b}$ should be exactly the same as $\frac{\partial^2 f}{\partial b\partial a}$ for the API \CodeIn{jax.lax.pow}. 
This inconsistency detected by \tech is confirmed and even labeled as ``\emph{urgent}'' by the \jax developers, which was fixed immediately after our report.

\subsection{Filtering Strategies}
\label{subsec:filter}

The gradient check for \tech checks gradients computed by totally different modes/implementations and may have more false positives than the output check (which checks values returned by largely shared implementations), as also confirmed by our result analysis in \S\ref{subsec:filter-rq}. Therefore, \tech further performs two additional filtering strategies for the gradient inconsistencies caused by numerical instability issues.%

\subsubsection{Differentiability}
\label{subsec:differentiability}

\begin{wrapfigure}{r}{0.45\columnwidth}
    \centering
    \includegraphics[keepaspectratio=true,width=0.45\columnwidth]{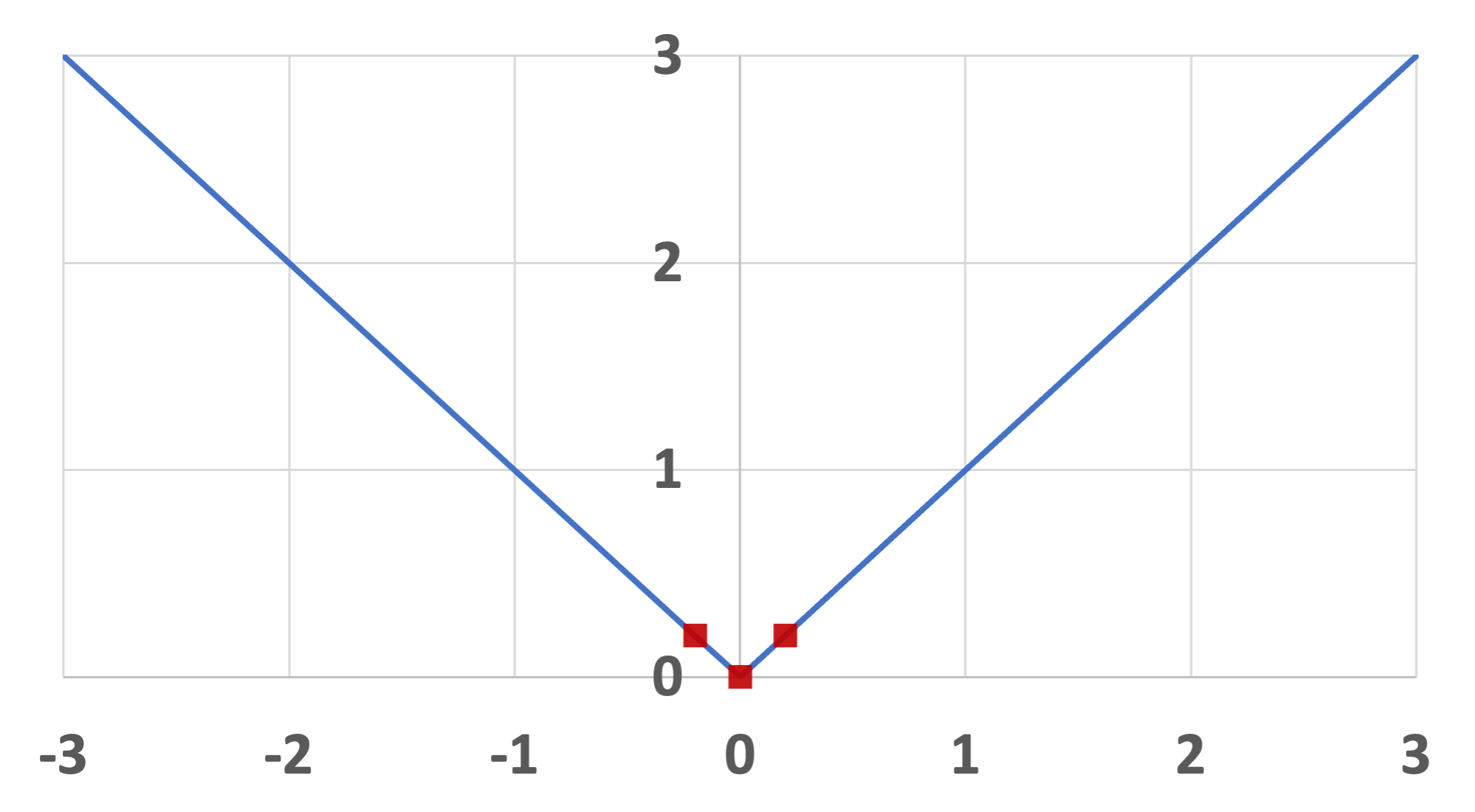}
    \caption{Abs function}
    \label{fig:abs}
\end{wrapfigure}

For the inconsistent gradients caused by non-differentiable points, we take the \textit{absolute} function as an example, which is shown in Figure~\ref{fig:abs}. 
Its gradient computed by \ad in \jax is 1 for input 0, but \nd will calculate the gradient as 0. 
However, this inconsistency is acceptable since the absolute function is non-differentiable at point 0. 
The gradient at the non-differentiable point is undefined, so \ad is allowed to return any value as the gradient.%
Thus, we need to filter out such cases caused by non-differentiable points. To do so, we first define the property of differentiability~\cite{enwiki:diff}:

\begin{definition} 
\label{def:diff}
\textbf{Differentiability.} A function $\vectorF$ is differentiable at $\pointx$ \textit{iff} 1) $\vectorF$ is continuous at $\pointx$, and 2) all partial derivatives of $\vectorF$ exist in the neighborhood of $\pointx$ and are continuous at $\pointx$.
\end{definition}

Based on Definition~\ref{def:diff}, to test the differentiability of a function $\vectorF$ at a point $\pointx$, we can sample some neighbors of $\pointx$. After sampling, the outputs and gradients at these points are computed and compared with the output and gradient at $\pointx$. 
Note that we choose \nd for differerntiability checking as 
the main test target is the \ad mechanism and we do not want to mistakenly treat \ad bugs as instability and miss them.

We will sample $N$ (default \numNeighbor) \textit{random} neighbors to check the differentiability.  
More specifically, we define \textit{random} neighbor as $\textit{random}(\pointx)=\pointx+\textit{uniform}(-\delta, +\delta)$, where sampling distance $\delta$ is a hyper-parameter (default 10$^{-4}$). 
If the output or gradient of any neighbor is different from the point $\pointx$, \tech will consider $\vectorF$ is non-differentiable at the point $\pointx$.%
Back to the absolute function example. 
For point 0, it is obvious that the gradient of its left neighbor is -1 and the gradient of its right neighbor is 1. 
Both of them are different from 0, which is the gradient computed by \nd at point 0. 
As a result, \tech will filter out this false-positive case.

\subsubsection{Precision Conversion}

\begin{figure}[t]
\includegraphics[keepaspectratio=true,width=1.0\columnwidth]{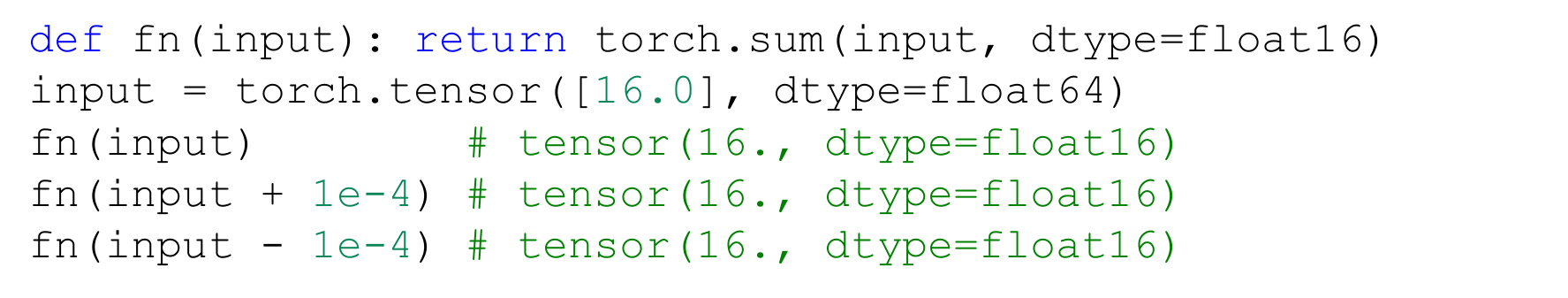}
	\caption{Example of precision loss}
	\label{fig:precision-conver}
\end{figure}

Figure~\ref{fig:precision-conver} shows a function \CodeIn{fn} which casts the \CodeIn{input} tensor to \CodeIn{float16} data type and returns the sum of all its elements~\cite{sum_website_pt}. 
Though \CodeIn{input} has data type \CodeIn{float64}, applying perturbation \CodeIn{1e-4} to it cannot change the output due to the loss of precision caused by rounding as shown in Figure~\ref{fig:precision-conver}. 
As a result, \nd will return 0 as the gradient, which is absolutely wrong. 
Besides, the precision conversion can also cause inconsistent gradients in reverse mode and forward mode \ad.%
In most DL libraries, the gradient computed by reverse mode \ad has the same data type as the input, while the gradient returned by forward mode has the same data type as the output. Thus, forward- and backward-mode \ad may have slight inconsistencies due to precision loss.
To filter out such inconsistencies, we exclude all cases where the API input and output have different precisions.

\section{Experimental Setup}

In the study, we address the following research questions:

\begin{itemize}
\item \textbf{RQ1:} Is \tech effective in detecting real-world bugs and improving code coverage?
\item \textbf{RQ2:} How do different components of \tech oracle affect its performance?
\item \textbf{RQ3:} How do the filter strategies contribute to the reduction of the false positive rate of \tech?
\end{itemize}

To answer the RQs, we have performed an extensive study on \pt, \tf, \jax, and \oneflow, whose details are shown in Table~\ref{tab:studiedLibrary}. 
With \numPtStars and \numTFStars stats on GitHub, \pt and \tf are the two most popular DL libraries, and they are also widely studied in prior DL library testing work~\cite{freefuzz,docter,eagle}. 
In addition, \jax~\cite{jax} and \oneflow~\cite{oneflow} are two emerging DL libraries, with \numJaxStars and \numOneflowStars stars on GitHub. 
\jax provides simple and powerful APIs for writing accelerated numerical code for high-performance machine learning research.%
With the growing research on training large models on distributed devices, \oneflow features a simple, neat redesign that enables easier programming of various parallelism paradigms compared to existing frameworks.

We compare \tech against both state-of-the-art model-level (\muffin~\cite{muffin}) and API-level (\freefuzz~\cite{freefuzz}) DL library fuzzers.
We run all experiments on a machine with 32-core AMD CPU (3.5GHz), 256GB RAM, and Ubuntu 20.04.

\begin{table}[!t]\centering
\caption{Details of the studied DL libraries}

\scalebox{0.85}{
\begin{tabular}{lrrrrr}
\toprule
& Github Stars  & Company  &\# Total APIs & \# Covered APIs & Version \\\midrule
\pt & \numPtStars & Meta & \numPtTotalAPI & \numPtCoverAPI & \verPt\\ 
\tf & \numTFStars &  Google &  \numTFTotalAPI & \numTFCoverAPI& \verTF \\
\jax & \numJaxStars &  Google & \numJaxTotalAPI & \numJaxCoverAPI & \verJax\\
\oneflow & \numOneflowStars &  OneFlow &  \numOneflowTotalAPI & \numOneflowCoverAPI & \verOneflow \\
\bottomrule
\end{tabular}}
\label{tab:studiedLibrary}
\end{table}

\subsection{Implementation}

\subsubsection{Input Generator} 
We leverage the input database and fuzzing strategies of \freefuzz~\cite{freefuzzrepo} to generate API inputs. While our approach is general, we choose \freefuzz since it is state-of-the-art and fully automated.%
We follow its default setting to generate (via mutation) 1000 inputs for each API. 
Because \freefuzz is only implemented for \pt and \tf, we further implement a \freefuzz-like fuzzing engine for \jax and \oneflow by ourselves. Following \freefuzz, we collect API inputs from open source and implement the fuzzing strategies. For the input collection of \jax, we only trace the developer tests (\numJaxTestFiles test files) since they already cover \ratioJaxCover \jax APIs (\numJaxCoverAPI/\numJaxTotalAPI). 
For \oneflow, we collect the input from all three sources: documentation, \numOneflowTestFiles developer tests, and \numOneflowModel DL models, covering 73.1\% (\numOneflowCoverAPI/\numOneflowTotalAPI) \oneflow APIs.

\subsubsection{Execution Scenarios}
\label{subsec:implementExe}

Table~\ref{tab:adapi} shows the example differentiation APIs used in our tool for each execution scenario. 
Note that not all the AD-related APIs we leverage are included in the table due to the space limit. 
The ``N/A'' in the table means the DL library does support or provide the API for that scenario. 
Only \oneflow has not implemented forward mode \ad, we thus skip the forward mode \ad testing for \oneflow.

For the DL libraries with APIs that could compare ND and AD gradients (shown in Column ``\nd''), \tech{} directly leverages such APIs. 
It turns out that only \oneflow does not have such an API, so we implement \nd for it by ourselves. 

\begin{table*}[!t]
\centering
\caption{Examples of \ad-related APIs of the studied DL libraries}
\scalebox{0.83}{
\begin{tabular}{lrrr}
\toprule
& Reverse Mode \ad & Forward Mode \ad & \nd\\ \midrule
\pt & \CodeIn{\footnotesize \ptRevAPI} & \CodeIn{\footnotesize \ptFwdAPI} & \CodeIn{\footnotesize \ptNDAPI} \\
\tf & \CodeIn{\footnotesize \tfRevAPI} & \CodeIn{\footnotesize \tfFwdAPI} & \CodeIn{\footnotesize \tfNDAPI} \\
\jax & \CodeIn{\footnotesize \jaxRevAPI} & \CodeIn{\footnotesize \jaxFwdAPI} & \CodeIn{\footnotesize \jaxNDAPI} \\
\oneflow & \CodeIn{\footnotesize \oneflowRevAPI} & N/A & N/A \\
\bottomrule
\end{tabular}
}
\label{tab:adapi}
\end{table*}

\subsubsection{Filter}
For the neighbor sampling of differentiability check, we set the sampling number $N$ as 5, and the distance $\delta$ as 10$^{-4}$ by default.
We also explore their impact in \S\ref{subsec:filter-rq}.

\subsection{Metrics}

\parabf{Number of Detected Bugs.} 
Following prior work on testing or fuzzing the DL libraries~\cite{freefuzz,muffin,docter,lemon,cradle,eagle, deeprel}, we report the number of bugs detected by \tech and compared baselines.

\parabf{False Positive Rate.} 
After filtering the inconsistent cases caused by instability, we get the bug candidates. 
However, not all candidates are real bugs.%
False positive rate (FPR) computes the proportion of the candidates that are false alarms, and is widely used in prior work on testing/fuzzing~\cite{su2021fully, you2019profuzzer, donaldson2017automated, deeprel}.
Following \muffin~\cite{muffin}, for every inconsistency reported by \tech, three authors independently inspected it to decide whether that is a bug or not and then discussed it together to reach a consensus. Moreover, different from \muffin, we further used developer feedback to calibrate our inspection. That said, any inconsistency will be reported as FP if the authors reach the consensus that this is not a bug or the developers rejected our report.

\parabf{Code Coverage.} 
Code coverage is one of the main criteria in software testing, and has also been recently adopted for testing DL libraries/compilers~\cite{freefuzz,muffin,liu2022coverageguided}. 
While state-of-the-art \freefuzz~\cite{freefuzz} and \muffin~\cite{muffin} only adopted \cpp or \python coverage, we adopt both the code coverage criteria for more thorough evaluation.%
Following \freefuzz and \muffin, we adopt line coverage, and trace the line coverage for \cpp and \python via GCOV~\cite{gcov} and \CodeIn{Coverage.py}~\cite{coverage-py}, respectively.

\parabf{Execution Time.} 
Since \tech leverages additional oracles for detecting \ad bugs, it would take more time than existing API-level fuzzers, such as \freefuzz. 
Thus, we take the execution time into account following prior work~\cite{freefuzz,lemon,liu2022coverageguided}. 

\section{Result Analysis}
\label{sec:result-analysis}

\subsection{RQ1: Detected Bugs and Coverage}
\label{subsec:rq1}

\subsubsection{Detected Bugs}

Table~\ref{tab:bugsummary} presents the summary of real-world bugs detected by \tech for all studied libraries. 
Column ``Total'' shows the total number of detected bugs. Column ``Confirmed (Fixed)'' presents the number of bugs confirmed and fixed by developers. 
We further categorize the confirmed bugs into previously unknown and known.
Plus, Column ``Rejected'' shows the number of bugs rejected by developers.
Lastly, Column ``Pending'' is the number of bugs not yet triaged by the developers. 

We can observe that \tech is capable of detecting \numTotalBugs bugs in total for the four studied DL libraries, with \numConfirmedBugs confirmed by developers and \numFixBugs already fixed, emphasizing the effectiveness of \tech. 
Notably, \numUnknownBugs are confirmed by developers as previously unknown bugs and only \numRejectBugs are rejected. Out of those \numConfirmedBugs confirmed bugs, state-of-the-art \freefuzz and \muffin can only detect \freefuzzDetect non-\ad bugs (all by \freefuzz and 0 by \muffin). 
{For these bugs detected by \freefuzz, 15 of them are unknown bugs and 6 are previously known. Of those unknown bugs, 10 are from JAX and OneFlow, the libraries not supported by the original FreeFuzz, and 5 are from PyTorch and TensorFlow.
6 bugs were rejected for the following reasons: 3 resulted from precision loss by using low-precision data types, 2 were intentionally implemented for numerical stability, and 1 arose from undefined behavior at a non-differentiable point.}

\begin{figure}[h]
\includegraphics[keepaspectratio=true,width=1.0\columnwidth]{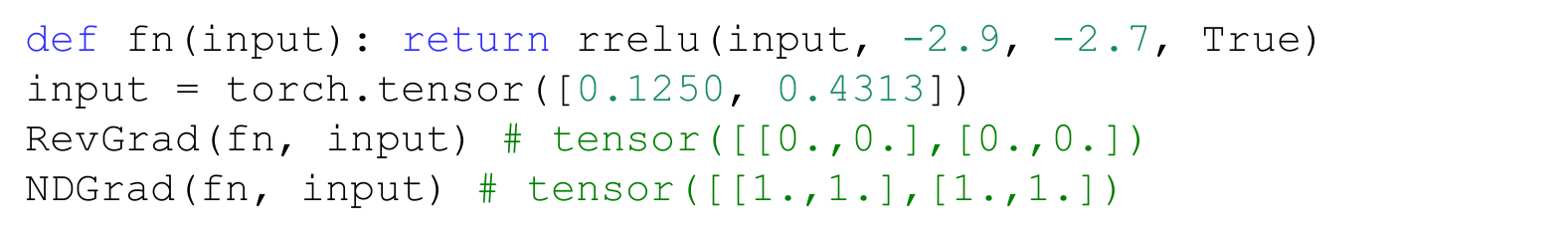}
	\caption{{High priority crash bug in \pt}}
	\label{fig:rrelu}
\end{figure}

{
Notably, 6 of our detected bugs for \pt are labeled with ``high-priority'' and 1 bug for \jax is labeled as ``P0(urgent)'' (all these 7 bugs are related to \ad) since they are critical and should be addressed urgently.
The other two libraries (\tf and \oneflow) do not have such labels so they are not discussed here.
Figure~\ref{fig:rrelu} shows a wrong gradient bug we detected in \CodeIn{rrelu}~\cite{rrelu_website_pt} which was commented by PyTorch developers as a \emph{massive} bug and labeled as ``high-priority'' and fixed immediately. 
For \pt, there are 78 high-priority bugs in total for its entire issue-tracking system during the two months of our issue reporting (May and June 2022), while 11 of them are related to \ad.
That said, \emph{\tech contributed 7.7\% of the high-priority bugs and 55.5\% for the high-priority \ad bugs}, showing the effectiveness of our approach.
The issue-tracking system of \jax has 22 ``urgent'' bugs in \emph{all-time} while only 1 of them is related to \ad, which is reported by us.
}

\begin{figure}[h]
\includegraphics[keepaspectratio=true,width=1.0\columnwidth]{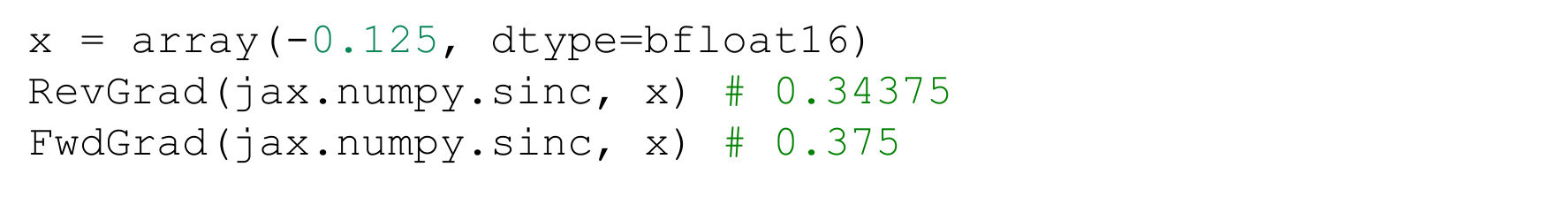}
	\caption{Inconsistent gradients in reverse/forward mode}
	\label{fig:jax-sinc-bug}
\end{figure}

Given multiple confirmed/fixed bugs have already been discussed in \S\ref{sec:approach}, here we will discuss an example rejected bug.
Figure~\ref{fig:jax-sinc-bug} shows an instance of \jax API \CodeIn{jax.numpy.sinc(x)}~\cite{sinc_website_jax}, which computes $\sin(\pi x)/(\pi x)$.
When the input \CodeIn{x} has the lowest precision floating datatype \CodeIn{bfloat16}~\cite{enwiki:bfloat16}, this API will have different gradients computed in forward mode and reverse mode \ad.
We reported this inconsistency to \jax developer, however, it was rejected: ``\emph{This is a consequence of the intended design of \CodeIn{bfloat16}. It is a worthwhile tradeoff for speed in deep learning contexts...}''.

\begin{table}[!t]
\centering
\caption{Summary of detected bugs}
\scalebox{0.85}{
\begin{tabular}{l|r|rr|r|r}
\toprule
& \multirow{2}{*}{\textbf{Total}} & \multicolumn{2}{c|}{\textbf{Confirmed (Fixed)}}  & \multirow{2}{*}{\textbf{Rejected}} & \multirow{2}{*}{{\textbf{Pending}}} \\ \cmidrule{3-4}
& & \textbf{Unknown} & \textbf{Known} & \\
\midrule
\pt & \numPtTotalBugs &  \numPtUnknownBugs (10) & \numPtKnownBugs (9) & \numPtRejectBugs 
 & {0} \\ 
\tf & \numTFTotalBugs & \numTFUnknownBugs (0) & \numTFKnownBugs (2)  & \numTFRejectBugs & {4} \\ 
\jax & \numJaxTotalBugs &  \numJaxUnknownBugs (5) & \numJaxKnownBugs (2) & \numJaxRejectBugs & {10} \\ 
\oneflow & \numOneflowTotalBugs & \numOneflowUnknownBugs (6) & \numOneflowKnownBugs (4) & \numOneflowRejectBugs & {9} \\ 
\midrule
\textbf{Total} & \numTotalBugs  & \numUnknownBugs (21)& \numKnownBugs (17) & \numRejectBugs & {23} \\
\bottomrule
\end{tabular}
}
\label{tab:bugsummary}
\end{table}

\subsubsection{Coverage}
\label{sebsec:rq1-cov}
We present the code coverage achieved by \tech and state-of-the-art \freefuzz on our default subjects, \pt and \tf, since they are not only the most popular DL libraries but also the only two libraries studied by \freefuzz. 
The comparison results on \jax/\oneflow are similar and omitted due to the space limit.
We follow the default setting of \freefuzz~\cite{freefuzz}, which executes 1000 mutated inputs for each API after running the seed inputs in the database.

\begin{figure}[t]
\centering

\begin{subfigure}[b]{0.49\columnwidth}
\centering
\includegraphics[keepaspectratio=true,width=0.95\textwidth]{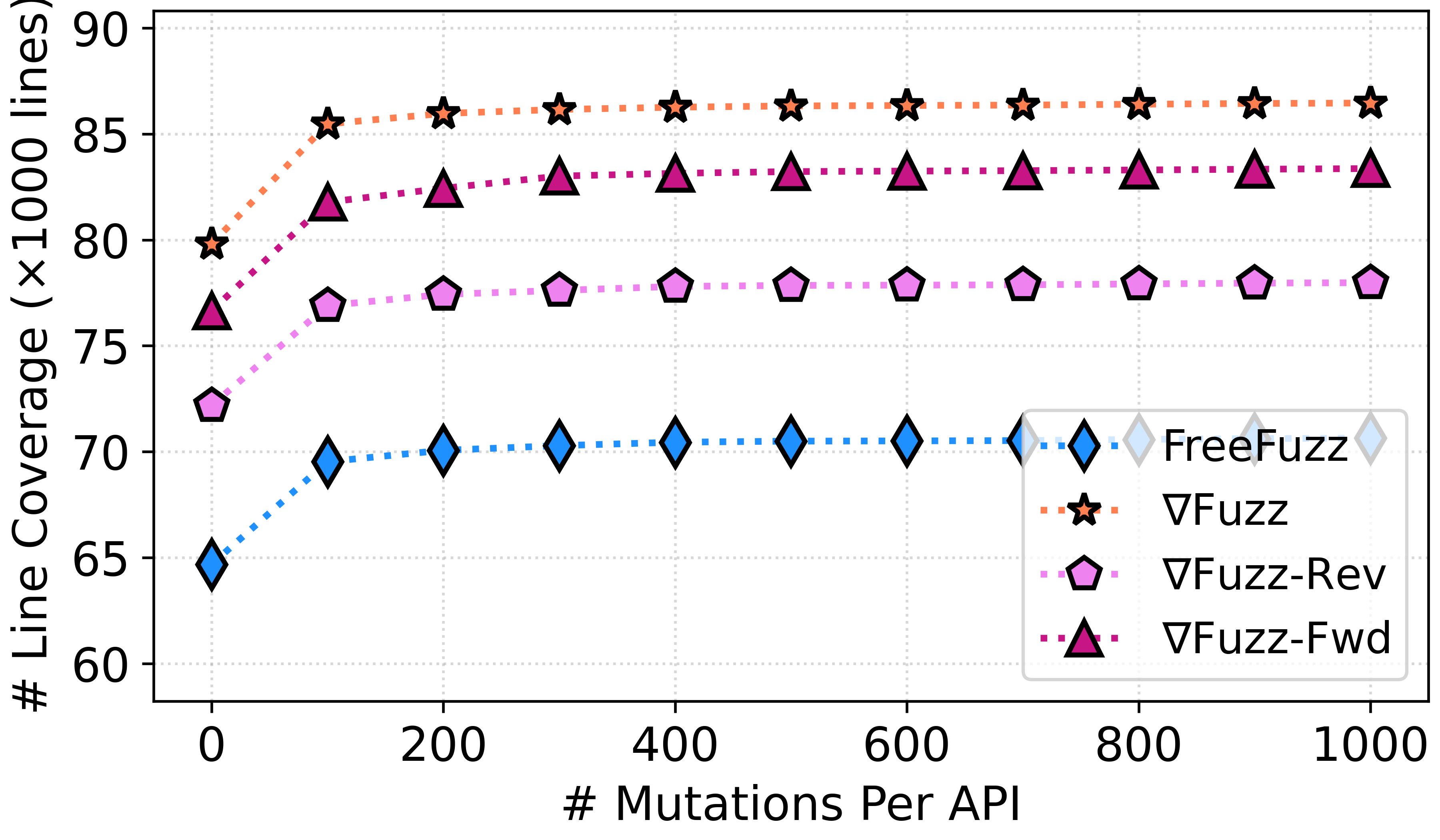}
\caption{\pt \cpp Coverage}
\label{fig:covptcpp}
\end{subfigure}
\begin{subfigure}[b]{0.49\columnwidth}
\centering
\includegraphics[keepaspectratio=true,width=0.95\textwidth]{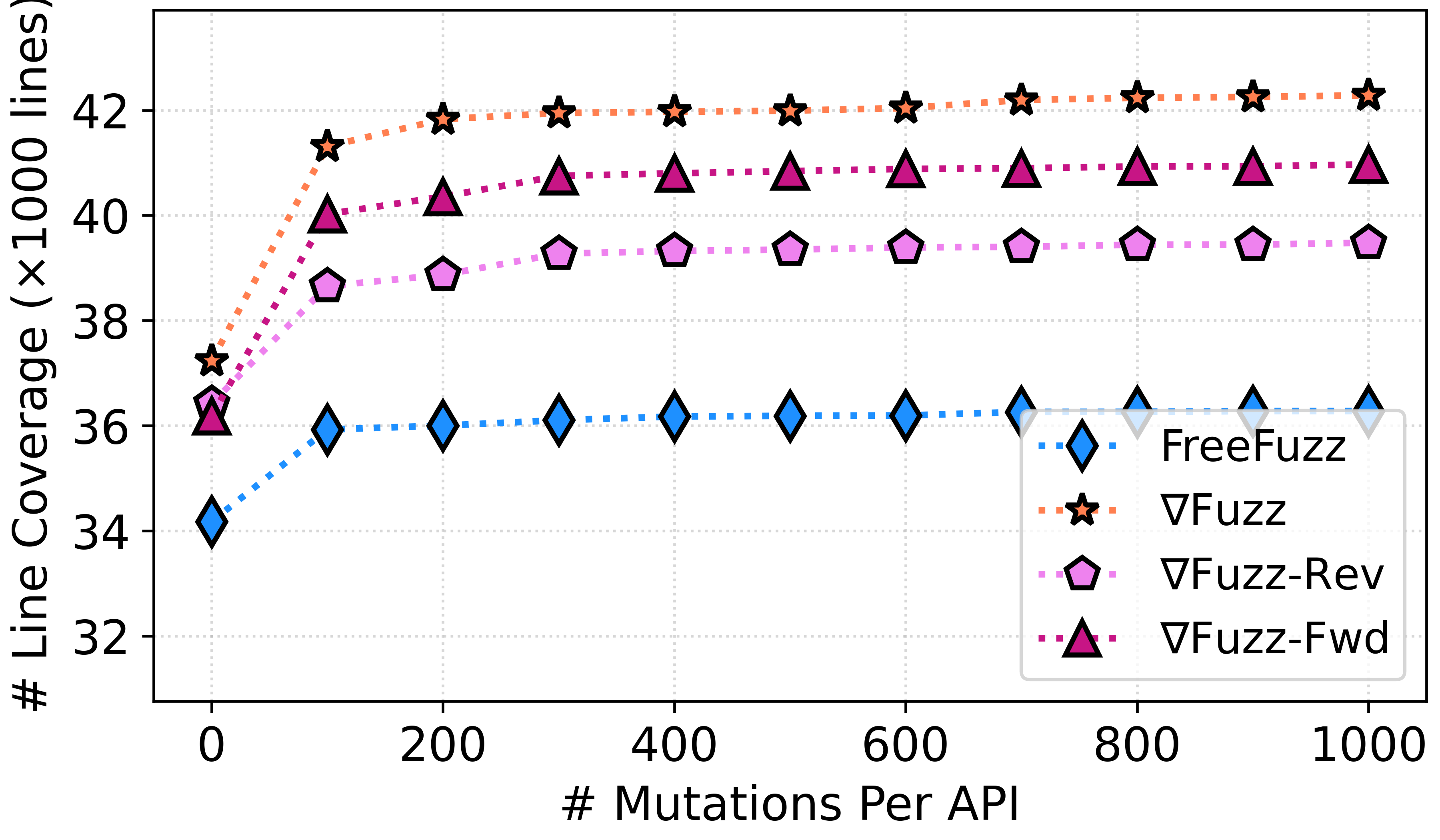}
\caption{\tf \cpp Coverage}
\label{fig:covtfcpp}
\end{subfigure}

\vfill

\begin{subfigure}[b]{0.49\columnwidth}
\centering
\includegraphics[keepaspectratio=true,width=0.95\textwidth]{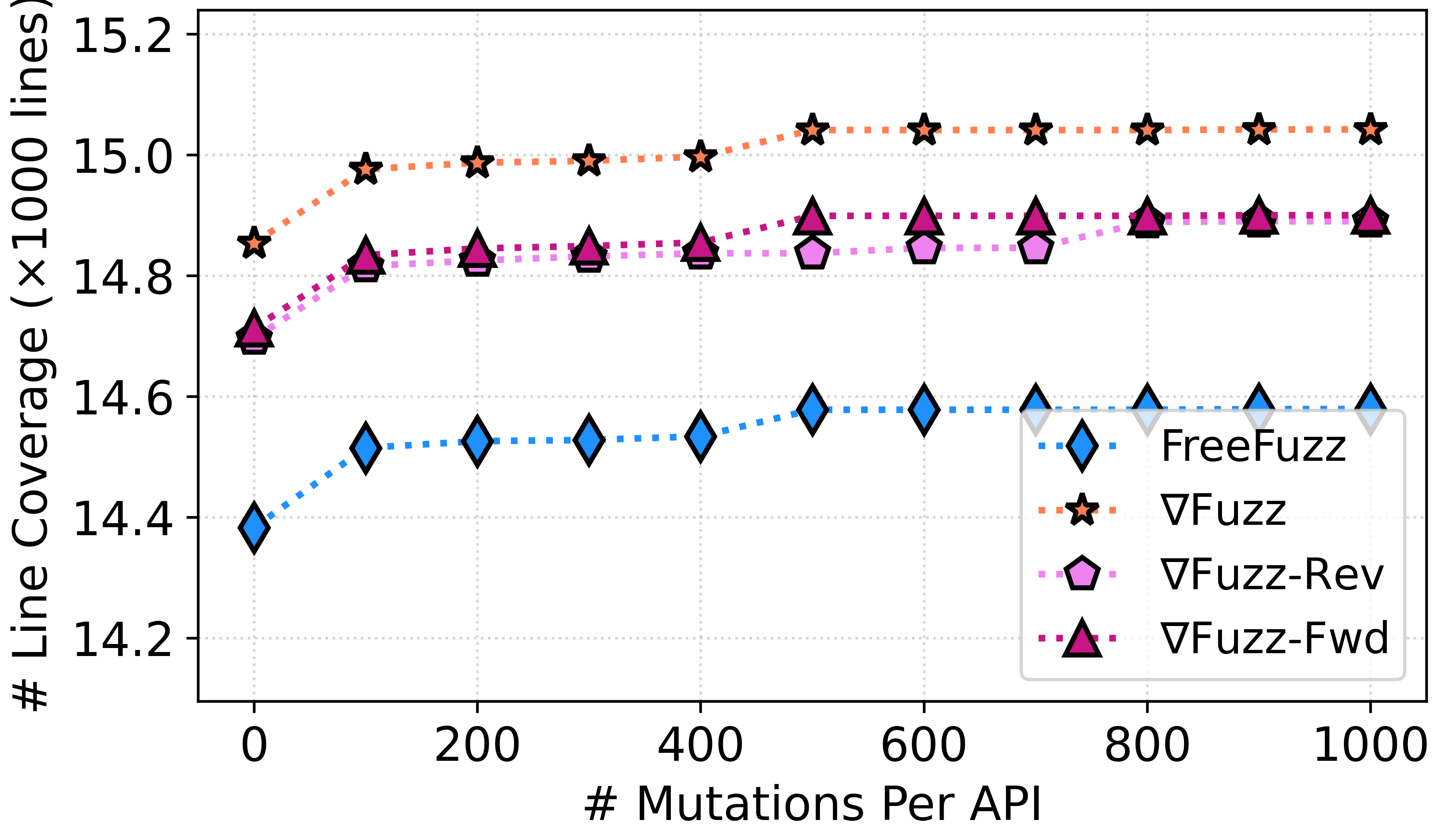}
\caption{\pt \python Coverage}
\label{fig:covptpy}
\end{subfigure}
\begin{subfigure}[b]{0.49\columnwidth}
\centering
\includegraphics[keepaspectratio=true,width=0.95\textwidth]{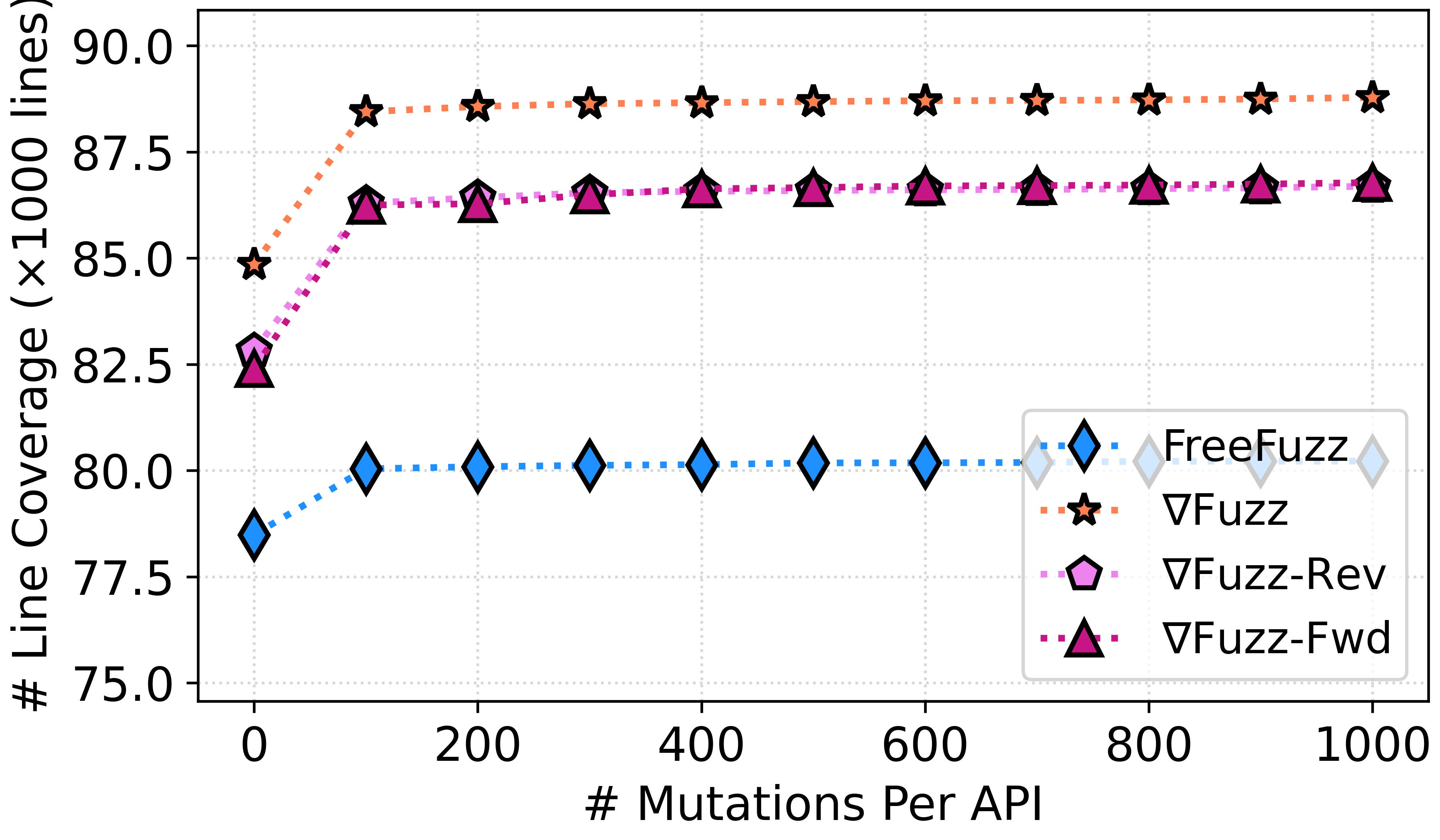}
\caption{\tf \python Coverage}
\label{fig:covtfpy}
\end{subfigure}

\caption{Coverage trend analysis}
\label{fig:covcpp}
\end{figure}

Figure~\ref{fig:covcpp} shows the coverage results, where the $x$ axis is the number of mutants generated for each API (from 100 to 1000 with the interval of 100), while the $y$ axis is the overall line coverage achieved. Note that the code coverage achieved by running the seed inputs in the \freefuzz database (without any mutation) is the start point for each line.%
For \cpp coverage, we can observe that \tech outperforms \freefuzz significantly on both \pt and \tf, with an improvement of 22.4\%/16.6\% respectively. Note that such an improvement is highly valuable as the additionally covered code is mostly about the crucial \ad mechanism.
For \python coverage, \tech still outperforms \freefuzz, but with a smaller improvement than \cpp. 
The possible reason could be that the crucial \ad functionality of DL libraries is mainly implemented in \cpp, e.g., the official material of \pt said, ``\emph{Autograd is a hotspot for PyTorch performance, so most of the heavy lifting is implemented in C++}''~\cite{autograd_pt}. 

Table~\ref{tab:freefuzz} further presents the time cost and overall system coverage rate for \freefuzz and \tech. 
The time cost of \tech is higher than \freefuzz due to the additional gradient computation (mostly on the expensive second-order gradients). Meanwhile, we can find that the \tech only running the seed inputs in the database (Row ``\tech (seed only)'') still outperforms \freefuzz in terms of code coverage even with less time. Moreover, \tech achieves decent system coverage rates, e.g., 25.8\% for the entire \pt \cpp codebase and 33.3\% for the entire \tf \python codebase.

We also compare \tech with \muffin. 
Since \muffin does not support \pt, \jax, or \oneflow, we conduct this comparison on \tf only. 
We run \muffin with its default setting (which takes 6.8h). For a fair comparison, we run \tech{} by setting the number of mutants for each API to 150, so it can finish within 6.8h. 
As shown in Table~\ref{tab:muffin}, with slightly less execution time (6.1h), \tech{} already substantially outperforms \muffin  in both code and API coverage.
In fact, even only running the seed API inputs without mutation with our \ad oracle (taking only 2.9h) is sufficient to outperform \muffin in terms of \cpp and \python coverage.
This is because \tech{} can cover much more APIs and more \ad modes, while \muffin only considers reverse mode \ad on a small set of APIs.
More precisely, \muffin only covers \numMuffinCoverAPI \tf APIs, while \tech can cover \numTFCoverAPI.
This is because \muffin only considers a set of predefined high-level layer APIs~\cite{muffin} for model generation. Meanwhile, \muffin can already achieve decent code coverage (albeit lower than \tech) because such high-level APIs will use various low-level operations.

\begin{table}[!t]
\centering
\caption{Comparison with \freefuzz}
\scalebox{0.8}{
\begin{tabular}{l|rrr|rrr}
\toprule
& \multicolumn{3}{c|}{\textbf{\pt}} & \multicolumn{3}{c}{\textbf{\tf}} \\ \cmidrule{2-7}
& \textbf{\cpp Cov} & \textbf{\python Cov}& \textbf{Time} & \textbf{\cpp Cov} & \textbf{\python Cov} & \textbf{Time} \\
\midrule
\freefuzz & \makecell[r]{70639\\(21.1\%)} & \makecell[r]{14579\\(13.9\%)} &  \freefuzzPtTimeCost & \makecell[r]{36279\\(9.77\%)} & \makecell[r]{80220\\(30.1\%)}& \freefuzzTfTimeCost\\
\midrule
\tech & \makecell[r]{86459\\(25.8\%)} & \makecell[r]{15042\\(14.3\%)} & \ptTimeCost & \makecell[r]{42284\\(11.4\%)}& \makecell[r]{88783\\(33.3\%)} & \tfTimeCost \\
\midrule
\makecell[l]{\tech\\(seed only)} & \makecell[r]{79808\\(23.4\%)} & \makecell[r]{14854\\(14.1\%)} & 1.4h & \makecell[r]{37233\\(10.0\%)} & \makecell[r]{84848\\(31.9\%)} & \tfTimeCostSeedOnly \\
\bottomrule
\end{tabular}
}
\label{tab:freefuzz}
\end{table}

\begin{table}[!t]
\centering
\caption{Comparison with \muffin }
\scalebox{0.85}{
\begin{tabular}{l|rrrr}
\toprule
 & \textbf{\cpp Coverage} & \textbf{\python Coverage} & \textbf{\# Covered API}  & \textbf{Time}\\
\midrule
\tech  & 41625 (11.21\%) & 88524 (33.24\%) & \numTFCoverAPI  & 6.1h\\
\muffin & 36884 (9.94\%) & 78754 (29.57\%) & \numMuffinCoverAPI  & 6.8h\\
\bottomrule
\end{tabular}}
\label{tab:muffin}
\end{table}

\subsection{RQ2: Different Components of Test Oracles}

\subsubsection{Impact on Bug Detection}

\begin{table}[!t]
\centering
\caption{Scenario distribution of confirmed bugs}
\scalebox{0.85}{
\begin{tabular}{l|r|rrr|r}
\toprule
& \multirow{2}{*}{\textbf{Direct}} & \multicolumn{3}{c|}{\textbf{\ad}} & \multirow{2}{*}{\textbf{\nd}}  \\ \cmidrule{3-5}
&\textbf{Invocation}  & \textbf{All} & \textbf{Rev-Only} & \textbf{Fwd-Only} & \\\midrule
\pt & \numPtDirect & \numPtAD & \numPtRev & \numPtFwd & \numPtND \\ 
\tf & \numTFDirect & \numTFAD& \numTFRev & \numTFFwd & \numTFND \\ 
\jax & \numJaxDirect & \numJaxAD & \numJaxRev & \numJaxFwd & \numJaxND \\ 
\oneflow & \numOneflowDirect & \numOneflowAD & \numOneflowRev & \numOneflowFwd & \numOneflowND \\ 
\midrule
\textbf{Total} & \numDirect & \numAD & \numRev & \numFwd & \numND \\
\bottomrule
\end{tabular}
}
\label{tab:bugdistribution}
\end{table}

In Table~\ref{tab:bugdistribution}, we categorize all confirmed bugs based on which execution scenarios they are located in, such as direct invocation, \ad, and \nd. 
Among the bugs in \ad, Column ``All'' displays the total number of bugs located in \ad, while Column ``Rev-Only''/``Fwd-Only'' presents the number of bugs \emph{only} appearing in reverse/forward mode respectively.%
That said, \numADMutual(\numAD-\numRev-\numFwd) bugs are in both reverse and forward AD modes.
From this table, we can conclude that most of the bugs detected by \tech are related to our main target \ad, showing the strength of \tech in fuzzing \ad for DL libraries.
One interesting fact is that we detect more bugs in direct invocation than AD in \oneflow. 
This may be because \oneflow was not tested by the original \freefuzz work and only supports reverse-mode \ad. 
Furthermore, we detect more reverse mode unique bugs than forward mode since reverse mode is more widely implemented in DL libraries.%
As mentioned in \S\ref{subsec:implementExe}, we directly leverage the \nd computation/comparison APIs in \pt, \tf and \jax. 
It turns out we can even detect \numND bugs in such APIs.

\begin{table}[!t]
\centering
\caption{Symptoms of confirmed bugs}
\scalebox{0.85}{
\begin{tabular}{l|r|rrr}
\toprule
 & \textbf{Output} &{\textbf{Gradient Total}} & \textbf{1st-order} & \textbf{2nd-order} \\\midrule
\pt & \numPtOutput & \numPtGrad & \numPtFirstGrad & \numPtSecondGrad \\ 
\tf & \numTFOutput & \numTFGrad & \numTFFirstGrad &\numTFSecondGrad \\ 
\jax & \numJaxOutput & \numJaxGrad & \numJaxFirstGrad & \numJaxSecondGrad \\ 
\oneflow & \numOneflowOutput & \numOneflowGrad & \numOneflowFirstGrad & \numOneflowSecondGrad \\ 
\midrule
\textbf{Total} & \numOutput & \numGrad & \numFirstGrad & \numSecondGrad \\
\bottomrule
\end{tabular}
}
\label{tab:bugsymptom}
\end{table}

We also categorize all the confirmed bugs by how they were detected in Table~\ref{tab:bugsymptom}, e.g., the bugs are found by inconsistent outputs (Column ``Output'') or gradients (Column ``Gradient Total''). 
We further split the bugs detected by the gradient into checks for the first-order gradients (Column ``1st-order'') and second-order gradients (Column ``2nd-order'').
We can observe that more than half bugs are detected by inconsistent gradients, showing the importance of gradient oracles. 
Plus, most of the gradient-related bugs are first-order. 
This is because the first- and second-order gradient computations often share part of the implementation, and any bug in the former will prevent \tech from testing the latter. 
Notably, \tech can still detect \numSecondGrad bugs using the second-order gradient check, showing the generality of our approach.
More interestingly, \numOutput bugs are revealed by discrepant outputs, indicating that the \ad mechanism could even affect normal DL API forward computation!

\subsubsection{Impact on Code Coverage}

To study the impact of execution scenarios on code coverage, we have two \tech variants: \tech-Rev (disabling reverse mode \ad) and \tech-Fwd (disabling forward mode \ad). 
We skip the coverage analysis of \nd since it is typically implemented based on the basic direct API invocations and can hardly cover new code.%
Figure~\ref{fig:covcpp} also presents the research findings for the studied variants with various amounts of mutations for each API. We can observe that the reverse mode \ad occupies a larger portion of the DL library implementation than the forward mode because \tech-Fwd outperforms \tech-Rev in terms of code coverage. 
This complies with the truth that reverse mode \ad is the main technique used in DL systems.%
More importantly, we can also observe that the code coverage of both reverse and forward mode \ad is not negligible, indicating the necessity of considering both of them for DL library fuzzing.

\subsection{RQ3: FPR and Effectiveness of the Filtering Strategies}
\label{subsec:filter-rq}

Table~\ref{tab:fpr} shows the results of the FPR, which is categorized based on the checks. 
Column ``Output''/``Gradient'' shows the FPR of output/gradient check respectively. 
Column ``Total'' is the overall FPR of our technique. 
Under Column ``Gradient'', Column ``All'' is the FPR when both filtering strategies are used, and Column ``N/A'' is the FPR without any strategy.  
Column ``Diff''/``Precision'' presents the FPR with only the differentiability/precision strategy, respectively. 
From the table, we can observe that the overall FPR of \tech is only around 20\%, implying the efficacy of \tech.
Notably, our filtering strategies do not remove any true bug mistakenly, as confirmed by our manual check for all reported inconsistencies. 
Besides, the FPR of the gradient oracle with filtering (\GradFPR) is much lower than without it (\GradNAFPR), showing the effectiveness of our filtering strategies.%
Also, we can observe that both filtering strategies are effective in reducing FPR. 
More precisely, the differentiability strategy is more helpful than the precision strategy, especially in \oneflow, where the latter cannot help reduce any false positives. 
Moreover, the FPR of the output oracle is lower than the gradient oracle (even after filtering). 
The main reason is that API outputs should not be affected by different \ad modes, while the computed gradients can be slightly different across reverse-/forward-mode \ad and \nd due to different underlying implementations.%

We further evaluate the impact of hyper-parameters, sampling number $N$ (default 5) and distance $\delta$ (default 10$^{-4}$).
The study is conducted on our default subjects, \pt and \tf, due to the space limit.%
For the impact of sampling number $N$, we run our experiments with different $N$ values of 1, 2, 5, 10 as shown in Table~\ref{tab:fprnumber}. 
The choice of $N$ contributes little to the FPR of gradient oracle since all of them are close. 
Plus, the FPR decreases as $N$ increases, as more neighbors to compare implies more false positives will be filtered. 
However, the time cost will also increase as $N$ increases (e.g. the time cost of $N$ = 10 is about 1.5X higher than that of $N$ = 5). 
Thus, $N$ = 5 is a trade-off between the FPR and the time cost.

As for the impact of distance $\delta$, we run with different $\delta$ values of 10$^{-1}$, 10$^{-2}$, 10$^{-3}$, 10$^{-4}$, 10$^{-5}$. 
The result is shown in Table~\ref{tab:fprdelta}. 
First, the FPR decreases as $\delta$ increases. 
This is because the neighbor with a farther distance is more likely to have a different gradient, causing some false positives to be filtered. 
However, it may also exclude the real bugs since the large distance may cause the gradient to change dramatically even at differentiable points. 
For example, in \pt, $\delta$ = 10$^{-3}$ could filter out 2 true positives. 
We choose $\delta$ = 10$^{-4}$ as our default setting since it does not filter out any true positives.

\subsection{Threats to Validity}

The main threat to internal validity lies in the implementation of \tech.
The authors have thoroughly tested and reviewed the code of \tech to lessen the threat. 
The threats to external validity mainly lie in the evaluation benchmarks used.
We evaluated \tech on four of the most widely-used and actively-maintained DL libraries to confirm the generality of our approach.
Moreover, we adopt detected bugs, code coverage, false positive analysis, and execution time to reduce the threats to construct validity for the metrics used.

\begin{table}[!t]
\centering
\caption{False positive rate (FPR)}
\scalebox{0.85}{
\begin{tabular}{l|r|rrrr|r}
\toprule
& \multirow{2}{*}{\textbf{Output}} & \multicolumn{4}{c|}{\textbf{Gradient}} & \multirow{2}{*}{\textbf{Total}}\\ \cmidrule{3-6}
&  & \textbf{All} & \textbf{Diff} & \textbf{Precision} & \textbf{N/A}  & \\\midrule
\pt & \ptOutputFPR & \ptGradFPR & \ptGradDiffFPR &\ptGradPreFPR & \ptGradNAFPR & \ptFPR \\ 
\tf & \tfOutputFPR & \tfGradFPR & \tfGradDiffFPR & \tfGradPreFPR & \tfGradNAFPR & \tfFPR\\ 
\jax & \jaxOutputFPR & \jaxGradFPR & \jaxGradDiffFPR & \jaxGradPreFPR & \jaxGradNAFPR & \jaxFPR \\ 
\oneflow & \oneflowOutputFPR & \oneflowGradFPR & \oneflowGradDiffFPR & \oneflowGradPreFPR & \oneflowGradNAFPR & \oneflowFPR \\ 
\midrule
\textbf{Total} & \OutputFPR & \GradFPR & \GradDiffFPR & \GradPreFPR & \GradNAFPR & \FPR \\
\bottomrule
\end{tabular}
}
\label{tab:fpr}
\end{table}

\begin{table}[!t]
\centering
\caption{FPR of gradient oracle w.r.t $N$}
\scalebox{0.85}{
\begin{tabular}{l|rrrr}
\toprule
\textbf{Sampling Number $N$}& \textbf{1} & \textbf{2} & \textbf{5} & \textbf{10} \\\midrule
\pt & 23.6\% & 22.6\% & \ptGradFPR & 21.2\% \\
\tf & 21.1\% & 21.1\% & \tfGradFPR & 21.1\% \\
\bottomrule
\end{tabular}
}
\label{tab:fprnumber}
\end{table}

\begin{table}[!t]
\centering
\caption{FPR of gradient oracle w.r.t $\delta$}
\scalebox{0.85}{
\begin{tabular}{l|rrrrrr}
\toprule
\textbf{Sampling Distance $\delta$} & $10^{-1}$ & $10^{-2}$ & $10^{-3}$ & $10^{-4}$ & $10^{-5}$ \\\midrule
\pt & 17.2\% & 19.3\% & 20.4\% & \ptGradFPR & 22.9\% \\
\tf & 7.1\% & 18.8\% & 18.8\% & \tfGradFPR & 21.1\%\\
\bottomrule
\end{tabular}
}
\label{tab:fprdelta}
\end{table}

\section{Related Work}

\cradle~\cite{cradle} is a pioneering work on DL library fuzzing, which
leverages differential testing to detect bugs by running existing DL models on different low-level DL libraries of \keras~\cite{keras}.
\audee~\cite{audee} and \lemon~\cite{lemon} further augment \cradle by applying search-based mutation strategies on existing DL models to cover more library code. While \lemon adopted advanced mutation rules (e.g., layer addition), it still only covers a small set of APIs~\cite{freefuzz} due to its strict mutation rules, e.g., an API cannot be added/removed in the model unless its input and output tensor shapes are identical.
More importantly, these techniques focus on the inference phase of DL models, and thus cannot detect any \ad bug. 
To mitigate this, the recent \muffin work~\cite{muffin} is proposed to detect bugs in both inference and training phases by generating DL models via a top-down approach. While \muffin can potentially cover reverse-mode \ad, it can only cover a small number of APIs in specific libraries, and cannot detect any confirmed \ad bug (please see \S\ref{sec:intro} for detailed discussion).
\revision{More recently, \nnsmith~\cite{nnsmith} leverages lightweight formal specifications to model each operator, and generates diverse and valid models via symbolic constraint solving. While \nnsmith has been demonstrated to be state-of-the-art model-level DL library fuzzer, it still only targets the inference phase, while our work is orthogonal and can be applied to further augment \nnsmith. %
}

Besides leveraging DL models for testing DL libraries, researchers have also investigated directly fuzzing DL library APIs. Meanwhile, DL library APIs are often exposed in Python, a dynamically typed language, making it hard even to determine the input types for DL APIs.%
To overcome this issue, \predoo~\cite{predoo} requires manually setting up API arguments for DL library fuzzing, and thus was only evaluated on 7 \tf APIs (due to the manual efforts). 
More recently, \docter~\cite{docter} synthesizes rules to extract API input constraints from DL library documentations, and then generates API inputs based on the constraints. However, it still requires manual efforts for annotating 30\% of API parameters. 
Different from above work, \freefuzz~\cite{freefuzz} is a fully-automated technique for DL library API fuzzing. More specifically, \freefuzz automatically tracks API inputs when running code mined from the open-source;
additional mutations are performed to generate more inputs based on tracked seed API inputs.
Another line of recent work designs other test oracles for DL libraries, e.g., \deeprel~\cite{deeprel} automatically infers relational APIs (e.g., the APIs that should return the same values/statuses when given the same inputs) as the oracle to detect inconsistency bugs for DL libraries, while \eagle~\cite{eagle} uses equivalent graphs to differentially test DL APIs.
While effective, none of the existing API-level techniques targeted the crucial \ad engines in DL libraries.

\revision{Different from the above model- and API-level fuzzers which struggle to cover valid API sequences for a large number of APIs (due to complicated input/shape constraints), the very recent LLMFuzz work~\cite{deng2022fuzzing} proposes to directly apply modern Large Language Models (LLMs)~\cite{codex} to generate diverse DL API sequences. The insight is that LLMs can implicitly learn intricate DL API constraints from DL programs in their massive training corpora. LLMFuzz demonstrates, for the first time, that modern LLMs (e.g., Codex~\cite{codex}) can be directly leveraged for end-to-end fuzzing of real-world systems. \tech is also orthogonal to LLMFuzz and can be further applied to enhance its oracle support.   }%

\section{Conclusion}

\tech is the first approach specifically targeting the AD engine in DL libraries, which is a crucial component of any DL system. 
It leverages different execution scenarios as test oracles to test first- and high-order gradients and incorporates an automated filter to reduce the false positives caused by numerical instability. 
The evaluation of \tech on \pt, \tf, \jax and \oneflow shows that \tech can detect \numTotalBugs bugs in total, with \numConfirmedBugs confirmed by developers (\numUnknownBugs of which are previously unknown) and \numFixBugs already fixed. 
Notably, \tech contributed 58.3\% (7/12) of all high-priority \ad bugs for \pt and \jax during a two-month period.

\parabf{Data Availability:}
Our code and data are available at~\cite{nablafuzzrepo}.

\section*{Acknowledgments}
 This work was partially supported by NSF grants CCF-2131943, and CCF-2141474.
We also acknowledge support from Google and Meta.

\onecolumn \begin{multicols}{2}
\bibliography{main}
\end{multicols}

\end{document}